\documentclass[aps,pra,twocolumn,showpacs,superscriptaddress,nofootinbib,10pt]{revtex4-2}

\usepackage[T1]{fontenc}
\usepackage{lmodern}
\usepackage[utf8]{inputenc}
\usepackage{microtype}
\usepackage{amssymb,amsfonts,amsmath,mathtools,bm,braket,graphicx,array,booktabs}
\usepackage{amsthm}
\usepackage[dvipsnames]{xcolor}
\usepackage[colorlinks=true,linkcolor=blue,citecolor=blue,urlcolor=blue]{hyperref}

\newtheorem{theorem}{Theorem}
\newtheorem{proposition}{Proposition}
\newtheorem{lemma}{Lemma}
\newtheorem{corollary}{Corollary}

\newcommand{\id}{\mathbb{I}}
\newcommand{\rank}{\operatorname{rank}}
\newcommand{\ran}{\operatorname{ran}}

\newcommand{\Tr}{\operatorname{Tr}}

\newcommand{\CC}{\mathbb{C}}

\newcommand{\Sep}{\operatorname{SEP}}
\newcommand{\conv}{\operatorname{conv}}

\begin{document}


\title{On the structure of low-rank PPT entangled states violating Peres conjecture}

\author{Gurvir Singh}
\affiliation{Department of Physical Sciences,
Indian Institute of Science Education and Research Mohali,
Knowledge City, Sector 81, SAS Nagar,
Punjab 140306, India}
\affiliation{Harish-Chandra Research Institute, Chhatnag Road, Jhunsi, Prayagraj - 211019, India.}

\author{Manik Banik}
\affiliation{S. N. Bose National Centre for Basic Sciences, Block JD, Sector III, Salt Lake, Kolkata 700 106, India}

\begin{abstract}
Bound-entangled states with positive partial transpose (PPT) can exhibit quantum steering and Bell nonlocality, disproving the long-standing Peres conjecture and its stronger steering counterpart. These counterexamples, obtained through explicit constructions or numerical searches, raise the question of whether they share a common structural origin. We address this question in the low-rank regime by asking whether known counterexamples can be obtained, up to invertible local filtering, from the complement of an unextendible product basis (UPB). We find that the answer depends crucially on the local dimension. In $3\otimes3$, the rank-four structure of PPT entangled states implies that both the steerable PPT state of Moroder \textit{et al.} and the Bell-nonlocal PPT state of V\'ertesi and Brunner can be obtained from UPB-complement states under invertible local filtering. For three qubits, in contrast, rank-four states that are PPT across every bipartition are separable across each cut. Moreover, comparison of the product-vector structure of the relevant kernels shows that the known three-qubit Bell-nonlocal PPT states cannot be obtained by invertible local filtering from the complement of any three-qubit UPB. Going beyond rank four, we further show that a known rank-eight three-qutrit state, supported on the complement of a UPB and PPT across every bipartition, is genuinely multipartite entangled. These results show that UPBs provide a unifying structural origin for important low-rank PPT entangled states in $3\otimes3$, but that nonlocal bound entanglement is not, in general, a UPB phenomenon.
\end{abstract}
\maketitle

\section{Introduction}

Since its recognition~\cite{Einstein1935,Schrodinger1935}, quantum entanglement has evolved from a foundational feature of quantum mechanics into a powerful information-theoretic resource, underpinning and enhancing a wide range of tasks in communication~\cite{Bennett1992,Bennett1993}, cryptography~\cite{Ekert1991}, computation~\cite{Briegel2009}, and metrology~\cite{Giovannetti2004}; for comprehensive reviews, see Refs.~\cite{Dowling2003,Deutsch2020,Huang2026}. Einstein, Podolsky, and Rosen (EPR) invoked precisely this feature to argue that the quantum-mechanical description of physical reality afforded by the wave function is incomplete~\cite{Einstein1935}. Nearly three decades later, John S. Bell recast the EPR premises into an empirically testable form~\cite{Bell1964}, showing that the correlations predicted by quantum theory cannot, in general, be reproduced by any model satisfying local causality~\cite{Bell1964,Bell1966,Mermin1993}. More recently, quantum information theory has sharpened this picture considerably, revealing a strict hierarchy of nonclassical correlations---entanglement, steering (a notion whose origins can be traced to Schrödinger's work~\cite{Schrodinger1935}), and Bell nonlocality~\cite{Wiseman2007,Horodecki2009,Brunner2014,Uola2020}.

Nonetheless, the interrelation among entanglement, steering, and Bell nonlocality is considerably more intricate than this hierarchy alone conveys. Werner exhibited bipartite entangled states admitting a local hidden-variable model for all projective measurements, and hence incapable of violating any Bell inequality in that measurement scenario~\cite{Werner1989}; Barrett later extended the construction to arbitrary local positive-operator-valued measures (POVMs)~\cite{Barrett2002}. The picture changes, however, once many copies are available: Masanes showed that, for bipartite states, distillability is equivalent to the possibility of violating the Clauser--Horne--Shimony--Holt (CHSH) inequality asymptotically~\cite{Masanes2006}. This equivalence sharpens the question to the undistillable regime, where it was long believed that entangled states with positive partial transpose (PPT)---necessarily bound entangled---admit local hidden-variable models, a conjecture attributed to Peres~\cite{Peres1999}. A stronger conjecture held that such states are not merely local but also unsteerable~\cite{Pusey2013,Skrzypczyk2014}. Both conjectures have since been overturned in their respective settings. The multipartite variant of the Peres conjecture fell first, with the discovery of multipartite bound-entangled states violating multipartite Bell inequalities~\cite{Dur2001,Vertesi2012}; the bipartite conjectures were subsequently challenged by explicit constructions demonstrating that PPT entangled states can exhibit steering~\cite{Moroder2014} and, remarkably, Bell nonlocality~\cite{Vertesi2014}. Thus, while the hierarchy itself remains intact, its boundaries are not determined by simple structural properties of the state, such as PPT-ness.

These counterexamples, bipartite and multipartite alike, establish that PPT entanglement does not by itself preclude steering or Bell nonlocality~\cite{Dur2001,Vertesi2012,Moroder2014,Vertesi2014}. What they do not reveal, however, is the structural origin of such states. Each was obtained either through an explicit construction or by numerical search, leaving open whether these examples are isolated specimens or manifestations of a common underlying structure. At low rank, this question becomes particularly tractable, since the range and kernel of a PPT entangled state are tightly constrained, and a series of results by Chen and {\DJ}okovi{'c} makes this structure precise~\cite{Chen2011a,Chen2011b,ChenDjokovic2012classes,Chen2013b}. In particular, for three qubits, rank-four PPT entangled states obey strong structural constraints across each bipartition, which in turn constrain the product vectors contained in their kernels~\cite{Chen2013b}. For two qutrits, the characterization is stronger still: every rank-four PPT entangled state arises, up to an invertible local transformation, from the complement of a minimal unextendible product basis (UPB)~\cite{Chen2011a,Chen2011b}. The complement of a UPB is characterized by a rigid product-vector structure in its kernel. This naturally raises the question of whether the known counterexamples are instances of this construction---that is, whether an invertible local filtering operation can carry the complement of some UPB to these states---or whether nonlocal bound entanglement possesses a structurally distinct origin. The role of invertible local filtering is particularly useful in this context because it preserves the product-vector geometry of the range and kernel. In particular, the induced bijection between product vectors in the kernels preserves structural features such as their cardinality and incidence structure. Thus, comparing the product-vector geometry of the kernels provides a direct test of whether two states can belong to the same invertible-local-filtering orbit, and hence whether a PPT-entangled state can originate from a UPB-complement state in this sense.

As we show, the answer depends crucially on the local dimension. For two qutrits, it is affirmative: the Chen--{\DJ}okovi{'c} characterization places every rank-four PPT entangled state in $3\otimes3$ within the UPB framework, so both the Moroder steering state and the V\'ertesi--Brunner Bell-nonlocal state have a UPB origin. For three qubits, in contrast, the situation is fundamentally different. We first observe that rank-four states that are PPT across every bipartition are separable across every cut, so that the relevant counterexamples must occur at higher rank. We then compare the product-vector geometry of the kernels of the known Bell-nonlocal states with that of the complement of a three-qubit UPB and show that no invertible local filtering connects the two. Going beyond rank four, we further consider a known rank-eight three-qutrit state supported on the complement of a UPB and PPT across every bipartition~\cite{KrishnanRanjanBanik2021}, and show from the structure of its support that it is genuinely multipartite entangled.

The remainder of the paper is organized as follows. Section~\ref{sec:prelim} fixes notation and recalls the necessary background on entanglement in multipartite quantum systems, Bell nonlocality (bipartite and multipartite) and Schrödinger steering, and the notion of UPBs; Section~\ref{sec:threequbit} treats the three-qubit case; Section~\ref{sec:twoqutrit} the two-qutrit case; and Section~\ref{sec:beyondrank4} the higher-rank multipartite setting, including the three-qubit rank restrictions and fully PPT-GME three-qutrit UPB constructions. We conclude in Section~\ref{sec:conclusion} with a discussion of open questions.

\section{Preliminaries}
\label{sec:prelim}
Throughout the manuscript, we use notation standard in the quantum information community. The state of a $d$-dimensional quantum system $S$, associated with the complex Hilbert space $\mathbb{C}^d_S$, is described by a density operator $\rho_S\in\mathcal{D}(\mathbb{C}^d_S)$, where $\mathcal{D}(\mathbb{C}^d_S)$ denotes the compact convex set of all density operators on $\mathbb{C}^d_S$. Whenever there is no ambiguity, we omit the system label for brevity. Every pure state, characterized by $\Tr(\rho^2)=1$, can be represented by a unit vector $\ket{\psi}\in\mathbb{C}^d$, with $\rho=\ket{\psi}\bra{\psi}$, where vectors differing only by a global phase represent the same physical state. Multipartite states are associated with the tensor product $\mathcal{H}\equiv\bigotimes_{\ell=1}^n\mathbb{C}^{d_\ell}_{A_\ell}$ of their constituent subsystem Hilbert spaces, i.e., $\rho_{A_1\ldots A_n}\in\mathcal{D}\big(\mathcal{H}\big)$, while a pure state is represented by $\ket{\psi}_{A_1\ldots A_n}\in\mathcal{H}$.

\subsection{Entanglement in multipartite systems}

A state of the form $\ket{\psi_{\text{fp}}}_{A_1\ldots A_n}=\bigotimes_{\ell=1}^n\ket{\psi^{(\ell)}}$, with $\ket{\psi^{(\ell)}}\in\mathbb{C}^{d_\ell}_{A_\ell}$, is called a fully product state, and probabilistic mixtures of such states are called fully separable states, i.e. 
\begin{align}
\rho_{\text{fs}}=\sum_{k}p_k\ket{\psi_{\text{fp}}^k}\bra{\psi_{\text{fp}}^k};\quad p_k\ge0~\&~\sum_kp_k=1.
\end{align}
The set of all fully separable states, denoted by $\mathrm{FSEP}$, forms a convex, compact, proper subset of $\mathcal{D}\big(\mathcal{H}\big)$.

A tripartite state $\rho_{ABC}$ is called separable across the $A|BC$ bipartition if it can be written as a probabilistic mixture of pure states that are product across this cut, i.e.,
\begin{align}
\rho_{ABC}=\sum_kp_k\ket{\psi^k}\bra{\psi^k}_A\otimes\ket{\psi^k}\bra{\psi^k}_{BC},
\end{align}
where, for all $k$, $\ket{\psi^k}_A\in\mathbb{C}^{d_A}$ and $\ket{\psi^k}_{BC}\in\mathbb{C}^{d_B}\otimes\mathbb{C}^{d_C}$. We denote the corresponding set by $\mathrm{SEP(A|BC)}$, which forms a convex, compact subset of $\mathcal{D}(\mathbb{C}^{d_A}\otimes\mathbb{C}^{d_B}\otimes\mathbb{C}^{d_C})$. A state $\rho_{ABC}$ is said to be PPT across the $A|BC$ bipartition if $\rho^{T_A}_{ABC}\ge0$, where $T_A$ denotes the partial transpose with respect to the subsystem $A$. The set $\mathrm{PPT(A|BC)}$ of states that are positive under partial transpose across $A|BC$ likewise forms a convex, compact subset. Analogous sets can be defined for the other bipartitions. For later use, we define the following important subsets:
\begin{subequations}
\begin{align}
\mathcal B_{\cap}
&:= \Sep(A|BC)\cap\Sep(B|AC)\notag\\
&\quad\cap\Sep(C|AB),\label{eq:bcap}\\
\mathcal B_{\rm ch}
&:= \conv\!\bigl[\Sep(A|BC)\cup\Sep(B|AC)\notag\\
&\quad\cup\Sep(C|AB)\bigr],\label{eq:bcup}\\
\mathcal P_{\cap}
&:= \operatorname{PPT}(A|BC)\cap\operatorname{PPT}(B|AC)\notag\\
&\quad\cap\operatorname{PPT}(C|AB),\label{eq:pcap}\\
\mathcal P_{\rm ch}
&:= \conv\!\bigl[\operatorname{PPT}(A|BC)\cup\operatorname{PPT}(B|AC)\notag\\
&\quad\cup\operatorname{PPT}(C|AB)\bigr].\label{eq:pcup}
\end{align}
\end{subequations}
States outside $\mathcal B_{\rm ch}$ are called genuinely multipartite entangled~\cite{Ghune2009}. Following Ref.~\cite{Jungnitsch2011}, states in $\mathcal P_{\rm ch}$ are called PPT mixtures. Importantly, membership in $\mathcal P_{\rm ch}$ can be characterized efficiently through semidefinite programming (SDP). Since every biseparable state is a PPT mixture, one has $\mathcal B_{\rm ch}\subseteq\mathcal P_{\rm ch}$, and hence every state outside $\mathcal P_{\rm ch}$ is necessarily genuinely multipartite entangled. For $(\mathbb{C}^d)^{\otimes3}$, one has the strict inclusion chain
\begin{equation}
\mathrm{FSEP}\subsetneq\mathcal B_{\cap}\subsetneq\mathcal B_{\rm ch}\subsetneq\mathcal P_{\rm ch},
\end{equation}
for all $d\ge2$. For the three-qubit case, the leftmost strict inclusion follows from the shift-UPB construction of Bennett \textit{et al.}~\cite{Bennett1999}. The state supported on the complementary subspace is separable across every bipartition and hence belongs to $\mathcal B_{\cap}$, while, by construction, it is not fully separable and therefore does not belong to $\mathrm{FSEP}$. The rightmost strict inclusion for three qubits was established in Ref.~\cite{Ha2016}, thereby disproving the conjecture of Ref.~\cite{Novo2013} that the sets $\mathcal B_{\rm ch}$ and $\mathcal P_{\rm ch}$ coincide.

\subsection{Bell nonlocality and Schrödinger steering}
\label{sec:bell}

{\it Bipartite scenario:} Bell nonlocality concerns the correlations generated by spatially separated parties performing local measurements on a shared quantum state. In a general bipartite Bell scenario, Alice and Bob choose measurement settings $x\in\mathcal X$ and $y\in\mathcal Y$, respectively, and obtain outcomes $a\in\mathcal A$ and $b\in\mathcal B$. A conditional probability distribution $p(ab|xy)$ is said to admit a local hidden-variable (LHV) model if it can be written in the form
\begin{equation}
p(ab|xy)=\int d\lambda~q(\lambda)p(a|x,\lambda)p(b|y,\lambda),
\label{eq:LHV}
\end{equation}
where $\lambda$ is a shared classical variable distributed according to $q(\lambda)$, and the response functions $p(a|x,\lambda)$ and $p(b|y,\lambda)$ depend only on the local measurement setting and the shared variable. Such correlations are called Bell local; correlations that do not admit a decomposition of the form~\eqref{eq:LHV} are called Bell nonlocal.

For a bipartite quantum state $\rho_{AB}$, local measurements are described by positive-operator-valued measures (POVMs) $\{M_{a|x}\}_a$ and $\{N_{b|y}\}_b$, satisfying $\sum_a M_{a|x}=\mathbb I_A$ and $\sum_b N_{b|y}=\mathbb I_B$. The resulting quantum correlation is given by Born's rule,
\begin{equation}
p(ab|xy)=\Tr\left[\rho_{AB}(M_{a|x}\otimes N_{b|y})\right].
\label{eq:quantumcorrelation}
\end{equation}
The state $\rho_{AB}$ is called Bell local if the correlations generated by every possible choice of local POVMs admit an LHV model; otherwise, it is called Bell nonlocal~\cite{Bell1964,Brunner2014}. 

{\it Schrödinger steering:} Quantum steering characterizes the ability of one party, through the choice of local measurements on a shared bipartite state, to remotely prepare different ensembles of states for the other party in a way that cannot be explained by a local-hidden-state (LHS) model~\cite{Schrodinger1935,Wiseman2007}. Let $\rho_{AB}$ be a bipartite state, and let Alice perform a POVM $\{M_{a|x}\}_a$ for measurement setting $x$. Bob's resulting (sub-normalized) conditional states, or assemblage, are
\begin{equation}
\sigma_{a|x}=\Tr_A\left[(M_{a|x}\otimes\mathbb I_B)\rho_{AB}\right].
\end{equation}
The assemblage is said to admit an LHS model if it can be written as
\begin{equation}
\sigma_{a|x}=\int d\lambda~q(\lambda)p(a|x,\lambda)\rho_\lambda,
\end{equation}
where $q(\lambda)$ is a probability distribution, $p(a|x,\lambda)$ is a local response function, and $\rho_\lambda$ is a quantum state of Bob's system that is independent of Alice's measurement choice $x$. A state $\rho_{AB}$ is called unsteerable from Alice to Bob if every assemblage generated by arbitrary local POVMs on Alice's side admits an LHS model; otherwise, $\rho_{AB}$ is called steerable from Alice to Bob. 

In the bipartite setting, Bell nonlocality forms the strongest level in the hierarchy
\begin{equation}
\text{Bell nonlocality}\subsetneq\text{EPR steering}\subsetneq\text{entanglement},
\end{equation}
where the inclusions refer to the corresponding sets of quantum states. Thus, every Bell-nonlocal state is steerable and every steerable state is entangled, while the converses do not hold in general~\cite{Wiseman2007,Brunner2014,Uola2020}. In particular, entanglement is necessary but not sufficient for Bell nonlocality.


{\it Multipartite nonlocality:} The notion of Bell nonlocality extends naturally to multipartite systems. For an $n$-partite state $\rho_{A_1\ldots A_n}$, local measurements performed by the parties generate conditional correlations $p(a_1\ldots a_n|x_1\ldots x_n)$, which are called multipartite Bell local if they admit a decomposition of the form
\begin{equation}
p(a_1\ldots a_n|x_1\ldots x_n)
=\int d\lambda~q(\lambda)\prod_{\ell=1}^n p(a_\ell|x_\ell,\lambda).
\end{equation}
Correlations that cannot be described in this form are multipartite Bell nonlocal and necessarily violate some multipartite Bell inequality~\cite{Brunner2014}. Multipartite nonlocality, however, admits a further distinction. A correlation may be nonlocal while being compatible with a convex mixture of correlations that are local across different bipartitions. Such correlations are referred to as biseparable or nongenuinely multipartite nonlocal. In contrast, a correlation is said to exhibit genuine multipartite Bell nonlocality if it cannot be expressed as a convex mixture of correlations that are Bell local across at least one bipartition. While the notion of genuine multipartite Bell nonlocality was introduced historically by Svetlichny~\cite{Svetlichny1987}, more operational formulations of genuine multipartite nonlocality were subsequently developed~\cite{Gallego2012,Bancal2013,Dutta2020}. Thus, genuine multipartite Bell nonlocality certifies nonlocal correlations that involve all parties in an intrinsically multipartite manner. This distinction parallels the hierarchy between biseparable and genuinely multipartite entangled states, but the two notions should not be conflated: genuine multipartite entanglement does not, in general, imply genuine multipartite Bell nonlocality~\cite{Brunner2014}.

\subsection{Unextendible product bases}
\label{sec:UPB}

The notion of an unextendible product basis (UPB) was introduced by Bennett \textit{et al.} in the study of bound entanglement~\cite{Bennett1999} (see also~\cite{DiVincenzo2003,Parthasarathy2004}). The central idea is that a subspace of a multipartite Hilbert space can contain a complete orthogonal set of product vectors that nevertheless cannot be extended by any additional product vector orthogonal to all of them. More precisely, consider a multipartite Hilbert space $\mathcal H=\bigotimes_{\ell=1}^n\mathbb C^{d_\ell}$. A set of mutually orthogonal product states
\begin{equation}
\mathcal U=\left\{\ket{\psi_k}=\bigotimes_{\ell=1}^n\ket{\psi_k^{(\ell)}}\right\}_{k=1}^m
\end{equation}
is called an unextendible product basis if there exists no nonzero product vector $\ket{\phi}=\bigotimes_{\ell=1}^n\ket{\phi^{(\ell)}}$ such that
\begin{equation}
\braket{\phi|\psi_k}=0,\qquad \forall,k=1,\ldots,m.
\end{equation}
An important aspect of UPBs is their connection to the phenomenon of `quantum nonlocality without entanglement'~\cite{Bennett1999NWE,Walgate2002,Halder2019}, where ensembles of mutually orthogonal product states, although locally preparable, cannot be perfectly distinguished by spatially separated parties using only local operations and classical communication (LOCC). This notion has subsequently been extended beyond quantum theory to the framework of generalized probabilistic theories (GPTs)~\cite{Bhattacharya2020NWE}. 

The another significance of UPBs stems from their direct connection to bound entanglement~\cite{Bennett1999}. Given a UPB, the normalized projector onto its orthogonal complement is entangled. At the same time, because the UPB consists of mutually orthogonal product states, $\rho_{\mathcal U}$ has positive partial transpose with respect to every subsystem, and hence provides a systematic construction of PPT bound-entangled states. Notably, UPBs are not unique to bipartite systems and admit a natural multipartite generalization~\cite{Bennett1999,DiVincenzo2003}. In mutipartite case, UPBs can also provide a systematic construction of genuinely entangled subspaces~\cite{Agrawal2019,Rout2019}. More specifically, let $\mathcal U=\{\ket{\psi_k}\}_{k=1}^m\subset\mathcal H$ be an UPB, and let the normalized projector onto its orthogonal complement is
\begin{equation}
\rho_{\mathcal U}=\frac{1}{D-m}\left(\mathbb I-\sum_{k=1}^m\ket{\psi_k}\bra{\psi_k}\right).
\label{eq:UPBstate}
\end{equation}
where $D:=\Pi_{\ell=1}^nd_\ell$ is the total dimension. Since the complement subspace $\mathcal U^\perp$ contains no fully product vector, the state $\rho_{\mathcal U}\notin\mathrm{FSEP}$, and hence is multipartite entangled. At the same time, because the UPB consists of mutually orthogonal product states, $\rho_{\mathcal U}$ has positive partial transpose with respect to every subsystem, and hence provides a systematic construction of PPT bound-entangled states. In the multipartite setting, however, the absence of fully product vectors from the support does not, by itself, determine the entanglement properties across the individual bipartitions. This distinction will be relevant later. 

Since one of our main questions is whether a given PPT-entangled state can be related to a UPB-complement state by invertible local filtering, it is useful to compare the structure of their ranges and kernels. We therefore record a simple observation concerning the action of invertible local operators. An invertible local operator on $\mathcal H=\bigotimes_{\ell=1}^n\mathbb C^{d_\ell}$ is of the form $X=\bigotimes_{\ell=1}^n X_\ell$, where each $X_\ell$ is invertible. Its action on a quantum state $\rho\in\mathcal{D}(\mathcal{H})$ is given by
\begin{equation}
\rho\longmapsto
\frac{X\rho X^\dagger}{\Tr(X\rho X^\dagger)}=\rho',
\label{eq:filtering}
\end{equation}
where we call $\rho'$ the filtered state obtained under local filtering $X$. The following elementary fact will be used throughout.

\begin{lemma}
\label{lem:invariant}
Let $\rho$ be a quantum state and let $X$ be an invertible local operator. Then, for the filtered state $\rho'=X\rho X^\dagger/\Tr(X\rho X^\dagger)$, one has
\begin{equation*}
\ker(\rho')=X^{-\dagger}\ker(\rho),\quad\ran(\rho')=X\ran(\rho),
\end{equation*}
where $X^{-\dagger}:=(X^\dagger)^{-1}$. Consequently, $X^{-\dagger}$ induces a bijection between the product rays contained in $\ker(\rho)$ and those contained in $\ker(\rho')$, while $X$ induces a bijection between the product rays contained in $\ran(\rho)$ and those contained in $\ran(\rho')$. The same statement holds for biproduct vectors across any fixed bipartition.
\end{lemma}

\begin{proof}
The normalization factor does not affect the range or kernel, so it suffices to consider $X\rho X^\dagger$. Since $X$ is invertible,
\begin{align*}
X\rho X^\dagger\ket{v}=0\Leftrightarrow X^\dagger\ket{v}\in\ker(\rho)\Leftrightarrow\ket{v}\in X^{-\dagger}\ker(\rho),
\end{align*}
which proves the kernel identity. For the range,
\begin{align*}
\ran(X\rho X^\dagger)=X\ran(\rho X^\dagger)=X\ran(\rho),
\end{align*}
where the second equality follows from the surjectivity of $X^\dagger$: $\ran(\rho X^\dagger)=\rho(\ran X^\dagger)=\rho(\mathcal H)=\ran(\rho)$. Finally, $X$, $X^{-1}$, $X^\dagger$, and $X^{-\dagger}$ are all tensor products of invertible local operators. Hence they map nonzero product vectors to nonzero product vectors in both directions. Regrouping the local factors according to any fixed bipartition gives the corresponding statement for biproduct vectors.
\end{proof}

Thus, invertible local filtering changes the explicit form of a state, but preserves the product-vector geometry of its range and kernel up to an invertible local transformation. In particular, the product-vector structure of the kernel is an invariant of the invertible-local-filtering orbit. We will use this observation to test whether known PPT-entangled counterexamples to the Peres conjecture  can be related to UPB-complement states by invertible local filtering.

\section{Three-qubit rank-four PPT entangled states}
\label{sec:threequbit}

Rank-four PPT entangled states of three qubits provide a natural setting in which to test whether Bell nonlocality can be traced to an underlying UPB structure. The reason is that two superficially similar rank-four constructions are available. The complement of the Shifts UPB gives a rank-four three-qubit state that is PPT across all bipartitions, while V\'ertesi and Brunner exhibited a rank-four, permutation-invariant, fully PPT three-qubit state that violates a tripartite Bell inequality~\cite{Vertesi2012}. The two states therefore agree at the level of rank and PPT properties. The question is whether they can nevertheless be distinguished by finer range--kernel geometry preserved by invertible local filtering.

The V\'ertesi--Brunner state, which we denote by $\varrho_{\rm VB}^{(3)}$, is invariant under partial transposition with respect to every qubit,
\begin{equation}
\left(\varrho_{\rm VB}^{(3)}\right)^{T_A}=\left(\varrho_{\rm VB}^{(3)}\right)^{T_B}=\left(\varrho_{\rm VB}^{(3)}\right)^{T_C}=\varrho_{\rm VB}^{(3)}.
\end{equation}
With the measurements given in Ref.~\cite{Vertesi2012}, it violates inequality number $5$ in \ifmmode\acute{S}\else'{S}\fi{}liwa's classification~\cite{Sliwa2003}, attaining $S\simeq3.0069$, above the local bound $S=3$~\cite{Vertesi2012}. Thus, the state is Bell nonlocal despite being separable across every bipartition. In particular, its Bell nonlocality is not genuine multipartite nonlocality in the sense of correlations that lie outside the convex set of bipartition-local models~\cite{Gallego2012,Bancal2013,Dutta2020}. The state was subsequently shown to provide an advantage in a multipartite communication-complexity task~\cite{Epping2013}.

Consider now thw three-qubit Shifts UPB~\cite{Bennett1999}:
\begin{equation}
\begin{aligned}
|\phi_1\rangle&=|0\rangle|1\rangle|+\rangle,
&|\phi_2\rangle&=|+\rangle|0\rangle|1\rangle,\\
|\phi_3\rangle&=|1\rangle|+\rangle|0\rangle,
&|\phi_4\rangle&=|-\rangle|-\rangle|-\rangle,
\end{aligned}
\end{equation}
where $|\pm\rangle=(|0\rangle\pm|1\rangle)/\sqrt2$. The corresponding UPB-complement state is
\begin{equation}
\rho_{\rm Shifts}
=
\frac14\left(\id-\sum_{i=1}^4|\phi_i\rangle\langle\phi_i|\right).
\label{eq:shiftsstate}
\end{equation}
It is again a rank-four state that is PPT across all three bipartitions. Thus, neither rank nor PPT properties alone distinguishes the two constructions. Before comparing their kernels, it is useful to isolate a consequence of rank four that is independent of UPBs. Each three-qubit bipartition is a $2\otimes4$ system, so the low-rank structure of PPT states becomes decisive.

\begin{proposition}[Rank-four cut separability]
\label{prop:rank4cuts}
Let $\rho$ be a three-qubit state with $\rank\rho\le4$ that is PPT across every bipartition. Then $\rho$ is separable across each of the three bipartitions, i.e., $\rho\in\mathcal B_{\cap}$. Consequently, $\rho$ is not genuinely multipartite entangled and cannot exhibit bipartite EPR steering across any fixed bipartition, in either direction.
\end{proposition}

\begin{proof}
Consider the bipartition $A|BC$, for which $\rho$ is a PPT state on $\mathbb C^2\otimes\mathbb C^4$ of rank at most four. The low-rank theorem for $2\otimes N$ systems states that a PPT state of rank at most $N$ is separable~\cite{Kraus2000}. Hence $\rho\in\mathrm{SEP(A|BC)}$. The same argument, after permuting the parties, gives $\rho\in\mathrm{SEP(B|AC)}
\qquad\text{and}\qquad
\rho\in\mathrm{SEP(C|AB)}$. Therefore $\rho\in\mathcal B_{\cap}$. Separability across any fixed bipartition also implies an LHS model for steering across that bipartition in both directions.
\end{proof}

Proposition~\ref{prop:rank4cuts} clarifies the status of the V\'ertesi--Brunner state from a complementary perspective. Its Bell violation is an ordinary tripartite Bell violation, whereas the state is separable across every bipartition. Thus it is neither genuinely multipartite entangled nor genuinely multipartite nonlocal under definitions that exclude convex mixtures of bipartition-local correlations~\cite{Gallego2012,Bancal2013,Dutta2020}. This proposition, however, does not resolve the structural question posed above, because the Shifts state also belongs to $\mathcal B_{\cap}$. We therefore need a finer invariant.

The required information is supplied by the rank-four structure theorem of Chen and {\DJ}okovi\'c~\cite{Chen2013b}. We state the part relevant to the present argument. A broader SLOCC classification of three-qubit rank-four PPT entangled states has also been developed recently~\cite{ChengChen2026}.

\begin{theorem}[Chen--{\DJ}okovi\'c~\cite{Chen2013b}]
\label{thm:CD3}
Let $\rho$ be a three-qubit PPT entangled state of rank four. Up to a permutation of the parties, it can be written as
\begin{equation}
\rho=\sum_{i=1}^4
|a_i\rangle\langle a_i|_A\otimes
|\psi_i\rangle\langle\psi_i|_{BC},
\end{equation}
where any two $|a_i\rangle$ are linearly independent, the four $|\psi_i\rangle$ are linearly independent, and each $|\psi_i\rangle$ is entangled. Let ${|\psi_i'\rangle}_{i=1}^4$ denote the reciprocal basis, defined by
$\langle\psi_i|\psi_j'\rangle=\delta_{ij}$. Then the kernel contains exactly four product rays across $A|BC$, namely
\begin{equation}
\left\{|a_i^\perp\rangle\otimes|\psi_i'\rangle,
\qquad i=1,\ldots,4\right\}.
\label{kernel_rays}
\end{equation}
\end{theorem}

The theorem reduces the search for fully product vectors in the four-dimensional kernel to four two-qubit problems. Since every fully product vector is, in particular, a product vector across $A|BC$, no additional $A|BC$ product ray can occur beyond those listed in Eq.~\eqref{kernel_rays}. This leads directly to a filtering invariant.

\begin{proposition}[Kernel product-vector count]
\label{prop:count}
Let $N(\rho)$ denote the number of fully product rays contained in the kernel of a rank-four three-qubit PPT entangled state. For the decomposition in Theorem~\ref{thm:CD3},
\begin{equation}
N(\rho)=\#\{i:C(\psi_i')=0\},
\end{equation}
where $C$ denotes the concurrence of a two-qubit pure state. Moreover, $N(\rho)$ is invariant under invertible local filtering.
\end{proposition}

\begin{proof}
By Theorem~\ref{thm:CD3}, every $A|BC$ product ray in $\ker \rho$ is proportional to one of the vectors in Eq.~(\ref{kernel_rays}). Such a vector is fully product when its $BC$ factor $|\psi_i'\rangle$ is a two-qubit product vector, which is equivalent to $C(\psi_i')=0$. The filtering invariance follows from Lemma~\ref{lem:invariant}, since an invertible local operator induces a bijection between the fully product rays contained in the two kernels.
\end{proof}

We can now apply the invariant directly, For the Shifts state all four reciprocal vectors are product and hence $N(\rho_{\rm Shifts})= 4.$ For the 
V\'ertesi--Brunner state, a direct evaluation gives the reciprocal-vector concurrences
\begin{equation}
    (0.1812, \; 0.2980, \; 0.2059, \; 1).
\end{equation}
Thus none of the kernel rays is fully product. The apparent similarity between the two states therefore disappears at the level of their kernel geometry.

\begin{corollary}
\label{cor:ceskernels}
The kernels of $\varrho_{\rm VB}^{(3)}$ contain no fully product vector, whereas the kernel of $\rho_{\rm Shifts}$ contains exactly four fully product rays.
\end{corollary}

This difference is the obstruction needed for the UPB question. Bravyi proved that every three-qubit orthogonal UPB contains four product vectors and that these are the only product vectors in its span~\cite{Bravyi2004}. The kernel of a three-qubit UPB-complement state is exactly this four dimensionsal UPB span, Combining this fact with Lemma~\ref{lem:invariant} and Corrollary~\ref{cor:ceskernels} gives the following conclusions. 

\begin{proposition}
\label{prop:notupb}
The V\'ertesi--Brunner state $\varrho_{\rm VB}^{(3)}$ cannot be obtained from a three-qubit UPB complement of a three-qubit UPB by by invertible local filtering.
\end{proposition}

\begin{proof}
Every three-qubit UPB complement has exactly four product vectors in its kernel, and every invertibly filtered representative has the same number by Lemma~\ref{lem:invariant}. Corollary~\ref{cor:ceskernels} gives $N(\varrho_{\rm VB}^{(3)})= 0.$
\end{proof}

Thus the first tripartite PPT counterexample to the peres conjecture is structurally different distinct from the three-qubit UPB complement family,
even though both constructions are rank four and fully PPT. This conclusion concerns invertible local filtering orbits; it does not assert tha every state in a three-qubit UPB-complement orbit is Bell local.

Another useful comparison is provided by the hidden-nonlocality construction of Tendick--Kampermann--Bru\ss~\cite{Tendick2020}. Their Bell-local parent state $\rho_L$ admits a local model for abritrary POVMs. After an invertible local filter produces, it is transformed into a state $\rho_{\rm NL}$ that violates the same \ifmmode\acute{S}\else\'{S}\fi{}liwa inequality considered above, 

\begin{equation}
\rho_L\xrightarrow{\text{invertible local filtering}}\rho_{\rm NL}.
\label{eq:tendick-filter}
\end{equation}

with the measurements of Ref.~\cite{Tendick2020} giving $S=3.0152$.

The four $A|BC$ product ryas in the kernel again give a convenient way to follow geometry through the filter. If $|\psi_{i,{\rm NL}}'\rangle$ denotes the $BC$ factor of one of the four kernel product rays of $\rho_{\rm NL}$, the corresponding factor for $\rho_L$ is, up to normalization,

\begin{equation}
|\psi_{i,L}'\rangle\propto
(G_B\otimes G_C)^{-\dagger}|\psi_{i,{\rm NL}}'\rangle,
\label{eq:reciprocal-filter}
\end{equation}
where $G_B$ and $G_C$ are the inverse local filters. Reconstructing the reciprocal vectors from the density matrix and filters reported in Ref.~\cite{Tendick2020} gives the values in Table~\ref{tab:conc}. The change along this filtering orbit is striking. For the Bell-local parent state, the four concurrences are of order $10^{-5}$--$10^{-4}$, close to the exactly vanishing values of the Shifts state, whereas after filtering they are of order $10^{-1}$--$1$. In this example reveleading hidden nonlocality is therefore accompanied by a substantial displacement of the reciprocal vectors away from the product-vector locus.

The tendick construction therefore gives a useful complement to the examples above. The number of fully product kernel rays is unchanged by invertible local filtering, but the geometry of the remaining kernel vectors can change substantially along the same filtering orbit. In the present case, the reciprocal vectors of the Bell-local parent lies very close to product vector locus, while those of the filtered Bell-nonlocal state are substantially more strongly entangled.

Taken together, these examples suggest that the relation between UPB structure and Bell-nonlocality is not captured by the rank or PPT properties alone. Exact product-vector content of the kernel provides a sharp distinction between UPB-complement states and V\'ertesi--Brunner construction, while the Tendick example shows that invertible local filtering can strongly reshape the geometry within a fixed filtering class. This motivates looking beyond the mere presence or absence of a UPB and asking how the geometry of the range and kernel constrains the nonclassical correlations that a PPT entangled state can display.

\begin{table}[t]
\centering
\scriptsize
\setlength{\tabcolsep}{2.6pt}
\begin{tabular}{@{}lcccc@{}}
\toprule
& $C(\psi'_1)$ & $C(\psi'_2)$ & $C(\psi'_3)$ & $C(\psi'_4)$\\
\midrule
$\rho_{\rm Shifts}$ & $0$ & $0$ & $0$ & $0$\\
V\'ertesi--Brunner & $0.1812$ & $0.2980$ & $0.2059$ & $1$\\
Tendick $\rho_L$ & $3.56\!\times\!10^{-4}$ & $1.43\!\times\!10^{-5}$ & $2.91\!\times\!10^{-4}$ & $3.82\!\times\!10^{-4}$\\
Tendick $\rho_{\rm NL}$ & $0.1324$ & $0.1468$ & $0.3616$ & $1$\\
\bottomrule
\end{tabular}
\caption{Concurrences of the reciprocal vectors associated with the four kernel product rays. The Tendick values are reconstructed from the rounded numerical state and filters reported in Ref.~\cite{Tendick2020}.}
\label{tab:conc}
\end{table}

\section{Rank-four two-qutrit states and their UPB structure}
\label{sec:twoqutrit}

The three-qubit result shows that a rank-four fully-PPT Bell-nonlocal need not have a UPB origin. As we will see, the two-qutrit case is qualitatively different. Here, rank four is the smallest rank at which PPT entanglement can occur, and the entire rank-four PPT entangled states admit a UPB characterization~\cite{Chen2011a,Chen2011b}. This makes it possible to answer the structural question without examining the kernel of each counterexample separately.

\begin{theorem}[Chen--{\DJ}okovi\'c~\cite{Chen2011b}]
\label{thm:CD33}
Let $\rho$ be a PPT state on $\CC^3\otimes\CC^3$ with $\rank\rho=4$. If $\rho$ is entangled, its range is a completely entangled subspace and there exist $A,B\in GL_3(\CC)$ together with a minimal five-state UPB $\{|u_k\rangle\}_{k=1}^5$ such that
\begin{equation}
\rho=
\frac{(A\otimes B)\Pi_{\rm UPB}(A^\dagger\otimes B^\dagger)}
{\Tr[(A\otimes B)\Pi_{\rm UPB}(A^\dagger\otimes B^\dagger)]}.
\label{eq:CD33form}
\end{equation}
where
$\Pi_{\rm UPB}:=\mathbb I_9-
\sum_{k=1}^{5}|u_k\rangle\langle u_k|$
is the orthogonal projector onto the
complement of the UPB subspace.\\
The kernel of $\rho$ contains precisely six product vectors in general position.
\end{theorem}

Thus the theorem says that every rank-four two-qutrit PPT entangled state has a minimal-UPB representative under invertible local filtering. It does not collapse all such states into a single SLOCC orbit; the inequivalent classes form a continuous family~\cite{ChenDjokovic2012classes}. For the present purpose, the important point is that the theorem applies simultaneously to the two PPT-entangled states that disproved the steering and Bell version of the Peres conjecture.

The PPT-invariant family used in the steering construction of Mororder \textit{et. al.}~\cite{Moroder2014} is
\begin{equation}
\label{eq:mor-family}
\varrho_{AB}
=\lambda_1|\chi_1\rangle\langle\chi_1|
+\lambda_2|\chi_2\rangle\langle\chi_2|
+\lambda_3\bigl(|\chi_3\rangle\langle\chi_3|+|\widetilde\chi_3\rangle\langle\widetilde\chi_3|\bigr),
\end{equation}
where
\begin{equation}
\label{eq:mor-states}
\begin{aligned}
|\chi_1\rangle&=\frac{|12\rangle+|21\rangle}{\sqrt2},\\
|\chi_2\rangle&=\frac{|00\rangle+|11\rangle-|22\rangle}{\sqrt3},\\
|\chi_3\rangle&=m_1|01\rangle+m_2|10\rangle+m_3(|11\rangle+|22\rangle),\\
|\widetilde\chi_3\rangle&=m_1|02\rangle-m_2|20\rangle+m_3(|21\rangle-|12\rangle).
\end{aligned}
\end{equation}
The parameters satisfy $m_i\ge0$ and $m_1^2+m_2^2+2m_3^2=1$. Invariance under partial transposition fixes
\begin{equation}
\label{eq:mor-lambdas}
\begin{gathered}
\lambda_3=\frac1\Delta,
\qquad
\lambda_1=1-\frac{2+3m_1m_2}{\Delta},\\
\lambda_2=1-\lambda_1-2\lambda_3,
\qquad
\Delta=4-2m_1^2+m_1m_2-2m_2^2.
\end{gathered}
\end{equation}
At $(m_1,m_2)=(0.2162,0.4363)$ the state violates the steering inequality used in Ref.~\cite{Moroder2014}.

The Bell-nonlocal two-qutrit state of V\'ertesi--Brunner~\cite{Vertesi2014} is another point of this same family, up to a local basis relabeling. The relevant parameter choice is
\begin{equation}
(m_1,m_2)=\left(\frac1{60},\frac3{10}\right).
\label{eq:VBpoint}
\end{equation}
Then
\begin{equation}
m_3=\frac{\sqrt{131/2}}{12},
\qquad
\Delta=\frac{1721}{450}.
\end{equation}
Substitution in Eq.~\eqref{eq:mor-lambdas} gives
\begin{equation}
\lambda=
\left(
\frac{3257}{6884},
\frac{450}{1721},
\frac{450}{1721},
\frac{27}{6884}
\right),
\label{eq:VBspectrum}
\end{equation}
which is exactly the spectrum of Ref.~\cite{Vertesi2014}. The same substitution gives the coefficient $\sqrt{131/2}/12$ that appears in its eigenvectors. The steering and Bell examples are therefore two parameter choices of one rank-four PPT family. The observation was already implicit in the note added to Ref.~\cite{Moroder2014}. The structural conclusion is now immediate, but it is useful to state it separately because it is the direct counterpart of Proposition~\ref{prop:notupb}.

\begin{corollary}
Both the Moroder \textit{et. al.}  steerable PPT state and the V\'ertesi--Brunner Bell-nonlocal PPT state are related by invertible local filtering to complements of states of two-qutrit UPBs.
\end{corollary}

\begin{proof}
Both states are entangled, PPT and rank four on $\mathbb{C}^3 \otimes \mathbb{C}^3$, so Theorem~\ref{thm:CD33} applies directly.
\end{proof}

It is important however to not identify this structural statement with Bell nonlocality of the particular orthogonal UPB-complement projector chosen as representative since Bell-nonlocality is not invariant under stochastic local filtering. Earlier locality results for UPB states naturally fit with this picture. Terhal \textit{et. al.} developed the symmetric-extension and symmetric-quasiextension method for constructing local hidden variable model in Bell-scenarios with fixed number of settings~\cite{Terhal2003}, and their examples included bound-entangled states built from UPBs. Thus UPB-based PPT entanglement already includes states with finite-setting locality properties. The Chen--{\DJ}okovi\'c theorem adds the complementary structural fact that Moroder and 
 V\'ertesi--Brunner counterexamples belong to the same broad UPB-generated rank-four family once invertible local filtering is allowed.

\section{Beyond rank four}
\label{sec:beyondrank4}
The rank-four analysis leaves open two different questions. First, can the UPB construction itself produce a fully-PPT GME state once we move away from the three-qubit setting? Second, if the answer is in negative for three-qubits, how tightly can one constrain a hypothetical three-qubit fully-PPT GME state? It will be useful to settle these points before turning to higher local dimension. 

For three qubits, the orthogonal-UPB route is closed. Bravyi's classification implies that every three-qubit orthogonal UPB contains four vectors~\cite{Bravyi2004}. Its normalized complement projector therefore has rank four and is PPT across all three bipartitions. Proposition~\ref{prop:rank4cuts} then forces it to be separable across each cut. Consequently, no three-qubit UPB-complement state can be genuinely multipartite entangled, even though it can fail to be fully separable. This obstruction is specific to the $2\otimes 4$ low-rank geometry rather than to UPBs in general.

The same low-rank theorem also restricts the broader three-qubit problem beyond UPB complements. In order to formulate the restriction, let us define the four ranks 

$$(r_0, r_A, r_B, r_C) = ( \rank \rho, \rank \rho^{T_A}, \rank \rho^{T_B}, \rank \rho^{T_C}).$$

Garberg \textit{et. al.} showed that an extremal state of the fully PPT three-qubit set must satisfy~\cite{Garberg2013}

\begin{equation}\label{eq:myrheim}
    r_0^2 + r_A^2 + r_B^2 + r_C^2 \leq 193.
\end{equation}

While this bound on its own does not say where a possible fully-PPT GME state should occur, it becomes useful after combining it with convexity and the rank-four obstruction.

\begin{corollary}[Rank window for a hypothetical three-qubit counterexample]
\label{cor:rankwindow}
If a three-qubit state exists that is PPT across all three bipartitions and genuinely multipartite entangled, then one may choose such a state to be extremal in $\mathcal P_{\cap}$ and to satisfy
\begin{equation}
    r_0, r_A, r_B, r_C \geq 5, \qquad \min\{r_0, r_A, r_B, r_C\} \leq 6.
\end{equation}
Thus at least one of the four ranks of an extremal counterexample must be either five or six.
\end{corollary}

\begin{proof}
Suppose $\rho_0 \in \mathcal P_{\cap} \setminus \mathcal B_{\rm ch}$. Since $\mathcal B_{\rm ch}$ is compact and convex, there is a Hermitian witness W separating $\rho_0$ from $\mathcal B_{\rm ch}$. Minimizing $\Tr(W\rho)$ over the compact convex set $\mathcal P_{\cap}$ produces an exposed face whose minimum is still negative. An extreme point of this face is therefore an extremal fully-PPT GME state. 
For such a state, $r_0 \leq 4$ is excluded by Proposition~\ref{prop:rank4cuts}.
If, for example, $r_A \leq 4$, then $ \sigma = \rho^{T_A} \geq 0$ and $\sigma^{T_A} = \rho \geq 0$, so $\sigma$ is a $2\otimes 4$ PPT state of rank at most four across $A|BC$. Hence $\sigma$ is separable across that cut by Ref.~\cite{Kraus2000}, and partial transposition maps the resulting product decomposition of $\rho$. The same argument excludes $r_B \leq 4$ and $r_C \leq 4$. Thus all four ranks are at least five. Finall, if all four were at least seven, the left-hand side of Eq.(\ref{eq:myrheim}) would be atleast $4\times 7^2 =196$, contradicting extremality. At least one is therefore five or six.
\end{proof}

Corollary~\ref{cor:rankwindow} however does not settle three-qubit existence problem. but it places any fully-PPT GME example beyond the rank-four UPB mechanism. Therefore any extremal example must occur at higher rank. To the best our knowledge, no explicit three-qubit state is presently known that is simultaneously PPT across all three bipartitions and genuinely multipartite entangled. 

The construction of Ha and Kye~\cite{HaKye2016} addresses a related but distinct question. Their states are genuinely tripartite entangled and PPT with respect to a chosen bipartition, and therefore provide PPT-mixture counterexamples to the conjecture that every three-qubit PPT mixture is biseparable. Full PPT requires the same state to satisfy 

\begin{equation}
    \rho^{T_A} \geq 0, \qquad \rho^{T_B} \geq 0, \qquad \rho^{T_C} \geq 0
\end{equation}
simultaneously, which is not required in that construction. 

The rank bounds also constrain the support of any hypothetical three-qubit example. Across each bipartition the support is then a subspace of $\mathbb{C}^2 \otimes \mathbb{C}^4$ of dimension at least five, whereas the largest completely entangled subspace in $\mathbb{C}^2 \otimes \mathbb{C}^4$ has dimension three~\cite{Parthasarathy2004}. 

Product vectors across each bipartition must therefore occur in the support. Genuine multipartite entanglement, if it occurs, cannot be explained simply by a product free support; it must depend on how the density operator occupies a support that already contains biproduct vectors. This makes the three-qubit problem a particularly clean setting in which to test the compatibility of full PPT with GME, genuine multipartite steering and genuine multipartite Bell nonlocality.

We now turn to three-qutrits, where the $\mathbb{C}^2 \otimes \mathbb{C}^4$ obstruction is absent. It was shown that
$\mathcal B_{\cap}\subsetneq\mathcal P_{\cap}$ in every tripartite local dimension~\cite{KrishnanRanjanBanik2021}. In the three-qutrit construction based on unextendible biseparable bases and genuinely entangled subspaces~\cite{Agrawal2019,KrishnanRanjanBanik2021}, let $\{|u_j\rangle\}_{j=1}^{19}$ denote the corresponding UPB and define
\begin{equation}
\rho^{[3]}_{(8)}
=
\frac18\left(
\id_{27}-\sum_{j=1}^{19}|u_j\rangle\langle u_j|
\right).
\label{eq:rho8}
\end{equation}
The state has rank eight and satisfies
\begin{equation}
(\rho^{[3]}_{(8)})^{T_A}
=(\rho^{[3]}_{(8)})^{T_B}
=(\rho^{[3]}_{(8)})^{T_C}
=\rho^{[3]}_{(8)}
\label{eq:rho8ppt}
\end{equation}
as follows from the real UPB construction~\cite{KrishnanRanjanBanik2021}. Inseparability across each bipartition alone does not imply genuine multipartite entanglement, because a biseparable state may use different cuts in different terms of a convex decomposition. We therefore examine the pure biseparable vectors contained in the support.

Define
\begin{equation}
d_k=|kkk\rangle,
\qquad k=0,1,2,
\end{equation}
and the six unnormalised face vectors
\begin{equation}
\begin{aligned}
f_1&=|0\rangle(|0\rangle+|1\rangle)(|1\rangle+|2\rangle),\\
f_2&=(|0\rangle+|1\rangle)|2\rangle(|1\rangle+|2\rangle),\\
f_3&=|2\rangle(|1\rangle+|2\rangle)(|0\rangle+|1\rangle),\\
f_4&=(|0\rangle+|1\rangle)(|1\rangle+|2\rangle)|0\rangle,\\
f_5&=(|1\rangle+|2\rangle)|0\rangle(|0\rangle+|1\rangle),\\
f_6&=(|1\rangle+|2\rangle)(|0\rangle+|1\rangle)|2\rangle.
\end{aligned}
\label{eq:faces}
\end{equation}
These nine vectors have disjoint computational supports. Their squared norms are $\|d_k\|^2=1$ and $\|f_\ell\|^2=4$. The range
\begin{equation}
\mathcal S=\ran\rho^{[3]}_{(8)}
\end{equation}
consists of the vectors
\begin{equation}
|\phi\rangle
=
a_0d_0+a_1d_1+a_2d_2+\sum_{\ell=1}^{6}b_\ell f_\ell
\label{eq:supportvector}
\end{equation}
subject to
\begin{equation}
a_0+a_1+a_2+4\sum_{\ell=1}^{6}b_\ell=0.
\label{eq:supportconstraint}
\end{equation}
The constraint follows because the remaining UPB vector in this nine-dimensional block is
$(|0\rangle+|1\rangle+|2\rangle)^{\otimes3}$.

Across $A|BC$, using the $BC$ ordering
$00,01,02,10,11,12,20,21,22$, the coefficient matrix of
$|\phi\rangle$ is
\begin{equation}
M_A=
\begin{pmatrix}
a_0&b_1&b_1&b_4&b_1&b_1&b_4&b_2&b_2\\
b_5&b_5&b_6&b_4&a_1&b_6&b_4&b_2&b_2\\
b_5&b_5&b_6&b_3&b_3&b_6&b_3&b_3&a_2
\end{pmatrix}.
\label{eq:MA}
\end{equation}
A vector is product across $A|BC$ exactly when $\rank M_A=1$. The following proposition identifies all such rays. Its role is to determine whether the biseparable vectors in the support can span the full rank-eight range.

\begin{proposition}
\label{prop:fourrays}
The only $A|BC$ product rays contained in $\mathcal S$ are
\begin{equation}
f_2-f_4,
\qquad
f_5-f_6,
\qquad
4d_0-f_1,
\qquad
4d_2-f_3.
\label{eq:fourrays}
\end{equation}
\end{proposition}

\begin{proof}
Let $R_0,R_1,R_2$ be the rows of $M_A$. We solve the rank-one condition by separating the cases $R_0\neq0$ and $R_0=0$.

Suppose first that $R_0\neq0$. Then there are numbers $r,s$ such that
$R_1=rR_0$ and $R_2=sR_0$. Comparing the entries of the three rows gives
\begin{align}
b_5&=r a_0=r b_1, & b_6&=r b_1, & a_1&=r b_1,
\label{eq:rankone-r1}\\
(1-r)b_4&=0, & (1-r)b_2&=0,
\label{eq:rankone-r2}
\end{align}
and
\begin{align}
b_5&=s a_0=s b_1, & b_6&=s b_1, & a_2&=s b_2,
\label{eq:rankone-s1}\\
b_3&=s b_1=s b_4=s b_2.
\label{eq:rankone-s2}
\end{align}
If $r=0$, then $R_1=0$, so
$b_5=b_6=b_4=b_2=a_1=0$. If $s\neq0$, Eq.~\eqref{eq:rankone-s1} with $b_5=0$ would give $a_0=b_1=0$, contradicting $R_0\neq0$. Hence $s=0$, and Eqs.~\eqref{eq:rankone-s1}--\eqref{eq:rankone-s2} give $b_3=a_2=0$. Only $a_0$ and $b_1$ remain, and Eq.~\eqref{eq:supportconstraint} gives $a_0=-4b_1$. This is the ray $4d_0-f_1$.

Now let $r\neq0$. Equation~\eqref{eq:rankone-r1} gives $a_0=b_1$. If also $b_1\neq0$, then Eqs.~\eqref{eq:rankone-r1} and~\eqref{eq:rankone-s1} give $s=r$. Since $b_3=s b_1=s b_4$, one has $b_4=b_1\neq0$, and Eq.~\eqref{eq:rankone-r2} therefore gives $r=1$. The same equations then give
\begin{equation}
a_0=a_1=a_2=b_1=b_2=b_3=b_4=b_5=b_6.
\end{equation}
Substitution into Eq.~\eqref{eq:supportconstraint} gives $27b_1=0$, contradicting $b_1\neq0$. Hence this branch has no nonzero solution.

It remains to take $r\neq0$ and $b_1=0$. Equation~\eqref{eq:rankone-r1} gives
$a_0=b_5=b_6=a_1=0$. Since $R_0\neq0$, at least one of $b_4,b_2$ is nonzero, and Eq.~\eqref{eq:rankone-r2} gives $r=1$. Equation~\eqref{eq:rankone-s2} gives $b_3=0$ and then $s=0$, while Eq.~\eqref{eq:rankone-s1} gives $a_2=0$. The support constraint reduces to $b_2+b_4=0$, which gives the ray $f_2-f_4$.

Finally suppose that $R_0=0$. Then
\begin{equation}
a_0=b_1=b_2=b_4=0.
\end{equation}
If $R_1=0$, then $b_5=b_6=a_1=0$, and Eq.~\eqref{eq:supportconstraint} leaves $a_2=-4b_3$, giving $4d_2-f_3$. If $R_1\neq0$, write $R_2=tR_1$. Comparing columns $4$, $7$, and $8$ gives $b_3=0$, and column $9$ gives $a_2=0$. If either $b_5$ or $b_6$ is nonzero, columns $1$--$3$ give $t=1$, while column $5$ then gives $a_1=0$. The support constraint becomes $b_5+b_6=0$, yielding $f_5-f_6$. The remaining possibility $b_5=b_6=0$ would require $a_1\neq0$ because $R_1\neq0$, but then columns $4$ and $5$ imply $t=0$ and hence $a_2=b_3=0$. Equation~\eqref{eq:supportconstraint} would force $a_1=0$, a contradiction.

The four rays in Eq.~\eqref{eq:fourrays} therefore exhaust all $A|BC$ product vectors in $\mathcal S$.
\end{proof}

The proposition gives more than inseparability across the three bipartitions.
Every vector in Eq.~\eqref{eq:fourrays} has zero $|111\rangle$ amplitude. The support is invariant under the cyclic permutation
\begin{equation}
C|a,b,c\rangle=|c,a,b\rangle.
\end{equation}
The cyclic images of Eq.~\eqref{eq:fourrays} therefore exhaust the product vectors across the other two bipartitions as well. Since $C|111\rangle=|111\rangle$, every pure biseparable vector contained in $\mathcal S$ lies in
\begin{equation}
\mathcal S\cap|111\rangle^\perp.
\label{eq:sevenplane}
\end{equation}
This is a proper seven-dimensional subspace of the eight-dimensional support. For example,
\begin{equation}
(|0\rangle+|1\rangle+|2\rangle)^{\otimes3}-27|111\rangle
\end{equation}
belongs to $\mathcal S$ and has nonzero $|111\rangle$ amplitude. Hence the biseparable pure states available inside the range cannot span the range itself. This is the missing ingredient needed to pass from entanglement across each individual bipartitions to GME.

\begin{theorem}
\label{thm:rho8gme}
The state $\rho^{[3]}_{(8)}$ is genuinely multipartite entangled. Together with Eq.~\eqref{eq:rho8ppt}, it is a fully-PPT GME three-qutrit state.
\end{theorem}

\begin{proof}
Suppose $\rho^{[3]}_{(8)}$ were biseparable. Refining a biseparable decomposition into pure states gives
\begin{equation}
\rho^{[3]}_{(8)}=\sum_i p_i|\phi_i\rangle\langle\phi_i|,
\end{equation}
where each $|\phi_i\rangle$ is product across at least one bipartition. Positivity of the decomposition implies $|\phi_i\rangle\in\mathcal S$, and the vectors $|\phi_i\rangle$ must span $\mathcal S$. Equation~\eqref{eq:sevenplane} places every such vector in a seven-dimensional subspace. This would imply $\rank\rho^{[3]}_{(8)}\le7$, contradicting $\rank\rho^{[3]}_{(8)}=8$.
\end{proof}

The earlier result places $\rho^{[3]}_{(8)}$ outside $\mathcal B_{\cap}$~\cite{KrishnanRanjanBanik2021}. This alone does not imply genuine multipartite entanglement. Theorem~\ref{thm:rho8gme} gives the stronger statement $\rho^{[3]}_{(8)}\notin\mathcal B_{\rm ch}$.

A recent construction of Shi \textit{et. al.} provides a complementary and particularly direct example~\cite{Shi2026}. They construct a minimum-cardinality genuinely unextendible product basis (GUPB) of fourteen states $\{|g_i\rangle\}_{i=1}^{14}$ in $(\mathbb{C}^3)^{\otimes 3}$. The normalized complement projector 

\begin{equation}
    \rho_{FS} = \frac{1}{13} \left (\mathbb{I}_{27} - \sum_{i=1}^{14} |g_i\rangle \langle g_i| \right)
\end{equation}

is again fully PPT and genuinely multipartite entangled. Since a GUPB is a UPB across every bipartition, so its orthogonal complement contains no biproduct vector at all; the range of $\rho_{FS}$ is therefore a genuinely entangled subspace. Moreover, the local vectors in the construction are real, giving

$$\rho_{FS}^{T_A} = \rho_{FS}^{T_B} = \rho_{FS}^{T_C} = \rho_{FS}$$

Thus the rank-eight state above and the rank-thirteen GUPB-complement state of Ref.~\cite{Shi2026} give two distinct UPB-based realizations of fully-PPT GME in three-qutrit case.

\section{Conclusion}
\label{sec:conclusion}

The low-rank analysis gives two rather different pictures. For three qubits, full PPT at rank four forces separability across every bipartition. The V\'ertesi--Brunner state can nevertheless violate a tripartite Bell inequality. Its kernel contains no fully product vector, while every three-qubit UPB-complement state has four kernel rays. Because this count is preserved by invertible local filtering, the V\'ertesi--Brunner state lies outside the three-qubit UPB-complement filtering family. 

The Tendick--Kampermann--Bru\ss\ construction adds a continuous version of same geometric viewpoint. For their state, the reciprocal-vector concurrences are 
of order $10^{-5}$--$10^{-4}$ for the Bell-local parent state and of order $10^{-1}$--$1$ after the filter that reveals Bell nonlocality. This makes the distance of the reciprocal vectors from the product locus a natural quantity to track in future searches for hidden nonlocality. 

For two qutrits, rank-four PPT entanglement is much more rigid. The Chen--{\DJ}okovi\'c theorem places every rank-four PPT entangled state within the UPB-generated family under invertible local filtering. The Moroder steering state and the V\'ertesi--Brunner Bell-nonlocal state are two members of the same PPT-invariant family. The symmetric-extension results of Terhal \textit{et. al.}~\cite{Terhal2003} show that UPB-based bound entanglement can also support local models in finite setting Bell scenarios. Taken together, these results make the role of the UPB construction structural rather than operational: it organizes the low-rank states, while steering and Bell behavior may vary among filtered representatives.

Beyond rank four, every three qubits orthogonal UPB still has four members, so its complement returns to the rank-four case. More generally,  Corollary~\ref{cor:rankwindow} shows that a fully-PPT GME three-qubit state, if one exists, can be chosen extremal with all four ranks at least five and atleast one $\rank \rho$, $\rank \rho^{T_A}$, $\rank \rho^{T_B}$ and $\rank \rho^{T_C}$  equal to five or six. The support of such a state must contain bi-product vectors across each bipartition, so any genuine multipartite entanglement would have to come from the mixed-state geometry rather than from a product-free support alone.

Three qutrits already allow explicit fully-PPT GME states. For the rank-eight state $\rho^{[3]}_{(8)}$, the bi-product vectors in its support are confined to a proper seven-dimensional subspace and therefore cannot span the range of a rank-eight biseparable state. The fourteen-state GUPB of Shi \textit{et. al.} gives a second construction whose complement contains no bi-product vectors across any bipartition~\cite{Shi2026}.

A natural question left open by these constructions is whether fully-PPT genuinely multipartite entangled states can also display genuinely multipartite forms of nonclassicality. In this setting, ordinary multipartite Bell nonlocality is not the appropriate benchmark. The three-qubit V\'ertesi--Brunner example already shows that a fully-PPT state may violate a tripartite Bell inequality while remaining separable across every bipartition. For a fully PPT GME state, the relevant question is instead whether the resulting correlations can lie outside the convex hull of correlations that are local, or nonsignalling, across some bipartition~\cite{Gallego2012,Bancal2013}. Likewise, one may ask whether such states exhibit genuine multipartite steering.

The three-qutrit UPB-complement states considered here provide concrete examples in which these questions can be investigated. They combine full PPT with genuine multipartite entanglement, thereby removing the obstruction present in the rank-four three-qubit examples. Determining whether this stronger entanglement structure can be converted into genuine multipartite steering or genuine multipartite Bell nonlocality would clarify how far the separation between PPT entanglement and nonclassicality when both the state and the observed correlations are required to be genuinely multipartite.

\begin{acknowledgments}
G.S. thanks Ritabrata Sengupta and Saronath Halder for helpful suggestions.\\
G.S. acknowledges financial support from the project “Technology Vertical - Quantum Communication” under the National Quantum Mission of the Department of Science and Technology (DST), Government of India (Sanction Order No. DST/QTC/NQM/QComm/2024/2(G)).
\end{acknowledgments}

\noindent
{\bf Declaration on the use of AI tools:} The research problem was conceived and the principal results were derived by the authors. In the course of this work, OpenAI’s ChatGPT (GPT-5.6 Sol) was used as an assistive tool for discussions concerning the proof of genuine multipartite entanglement of the three-qutrit PPT state. The authors have independently verified the arguments and take full responsibility for their correctness.


\begin{thebibliography}{59}%
\makeatletter
\providecommand \@ifxundefined [1]{%
 \@ifx{#1\undefined}
}%
\providecommand \@ifnum [1]{%
 \ifnum #1\expandafter \@firstoftwo
 \else \expandafter \@secondoftwo
 \fi
}%
\providecommand \@ifx [1]{%
 \ifx #1\expandafter \@firstoftwo
 \else \expandafter \@secondoftwo
 \fi
}%
\providecommand \natexlab [1]{#1}%
\providecommand \enquote  [1]{``#1''}%
\providecommand \bibnamefont  [1]{#1}%
\providecommand \bibfnamefont [1]{#1}%
\providecommand \citenamefont [1]{#1}%
\providecommand \href@noop [0]{\@secondoftwo}%
\providecommand \href [0]{\begingroup \@sanitize@url \@href}%
\providecommand \@href[1]{\@@startlink{#1}\@@href}%
\providecommand \@@href[1]{\endgroup#1\@@endlink}%
\providecommand \@sanitize@url [0]{\catcode `\\12\catcode `\$12\catcode
  `\&12\catcode `\#12\catcode `\^12\catcode `\_12\catcode `\%12\relax}%
\providecommand \@@startlink[1]{}%
\providecommand \@@endlink[0]{}%
\providecommand \url  [0]{\begingroup\@sanitize@url \@url }%
\providecommand \@url [1]{\endgroup\@href {#1}{\urlprefix }}%
\providecommand \urlprefix  [0]{URL }%
\providecommand \Eprint [0]{\href }%
\providecommand \doibase [0]{https://doi.org/}%
\providecommand \selectlanguage [0]{\@gobble}%
\providecommand \bibinfo  [0]{\@secondoftwo}%
\providecommand \bibfield  [0]{\@secondoftwo}%
\providecommand \translation [1]{[#1]}%
\providecommand \BibitemOpen [0]{}%
\providecommand \bibitemStop [0]{}%
\providecommand \bibitemNoStop [0]{.\EOS\space}%
\providecommand \EOS [0]{\spacefactor3000\relax}%
\providecommand \BibitemShut  [1]{\csname bibitem#1\endcsname}%
\let\auto@bib@innerbib\@empty
\bibitem [{\citenamefont {Einstein}\ \emph {et~al.}(1935)\citenamefont
  {Einstein}, \citenamefont {Podolsky},\ and\ \citenamefont
  {Rosen}}]{Einstein1935}%
  \BibitemOpen
  \bibfield  {author} {\bibinfo {author} {\bibfnamefont {A.}~\bibnamefont
  {Einstein}}, \bibinfo {author} {\bibfnamefont {B.}~\bibnamefont {Podolsky}},\
  and\ \bibinfo {author} {\bibfnamefont {N.}~\bibnamefont {Rosen}},\ }\bibfield
   {title} {\bibinfo {title} {{Can Quantum-Mechanical Description of Physical
  Reality Be Considered Complete?}},\ }\href
  {https://doi.org/10.1103/PhysRev.47.777} {\bibfield  {journal} {\bibinfo
  {journal} {Phys. Rev.}\ }\textbf {\bibinfo {volume} {47}},\ \bibinfo {pages}
  {777} (\bibinfo {year} {1935})}\BibitemShut {NoStop}%
\bibitem [{\citenamefont {Schr{\"o}dinger}(1935)}]{Schrodinger1935}%
  \BibitemOpen
  \bibfield  {author} {\bibinfo {author} {\bibfnamefont {E.}~\bibnamefont
  {Schr{\"o}dinger}},\ }\bibfield  {title} {\bibinfo {title} {{Discussion of
  Probability Relations between Separated Systems}},\ }\href
  {https://doi.org/10.1017/S0305004100013554} {\bibfield  {journal} {\bibinfo
  {journal} {Math. Proc. Camb. Philos. Soc.}\ }\textbf {\bibinfo {volume}
  {31}},\ \bibinfo {pages} {555} (\bibinfo {year} {1935})}\BibitemShut
  {NoStop}%
\bibitem [{\citenamefont {Bennett}\ and\ \citenamefont
  {Wiesner}(1992)}]{Bennett1992}%
  \BibitemOpen
  \bibfield  {author} {\bibinfo {author} {\bibfnamefont {C.~H.}\ \bibnamefont
  {Bennett}}\ and\ \bibinfo {author} {\bibfnamefont {S.~J.}\ \bibnamefont
  {Wiesner}},\ }\bibfield  {title} {\bibinfo {title} {{Communication via One-
  and Two-Particle Operators on Einstein-Podolsky-Rosen States}},\ }\href
  {https://doi.org/10.1103/PhysRevLett.69.2881} {\bibfield  {journal} {\bibinfo
   {journal} {Phys. Rev. Lett.}\ }\textbf {\bibinfo {volume} {69}},\ \bibinfo
  {pages} {2881} (\bibinfo {year} {1992})}\BibitemShut {NoStop}%
\bibitem [{\citenamefont {Bennett}\ \emph {et~al.}(1993)\citenamefont
  {Bennett}, \citenamefont {Brassard}, \citenamefont {Cr{\'e}peau},
  \citenamefont {Jozsa}, \citenamefont {Peres},\ and\ \citenamefont
  {Wootters}}]{Bennett1993}%
  \BibitemOpen
  \bibfield  {author} {\bibinfo {author} {\bibfnamefont {C.~H.}\ \bibnamefont
  {Bennett}}, \bibinfo {author} {\bibfnamefont {G.}~\bibnamefont {Brassard}},
  \bibinfo {author} {\bibfnamefont {C.}~\bibnamefont {Cr{\'e}peau}}, \bibinfo
  {author} {\bibfnamefont {R.}~\bibnamefont {Jozsa}}, \bibinfo {author}
  {\bibfnamefont {A.}~\bibnamefont {Peres}},\ and\ \bibinfo {author}
  {\bibfnamefont {W.~K.}\ \bibnamefont {Wootters}},\ }\bibfield  {title}
  {\bibinfo {title} {{Teleporting an Unknown Quantum State via Dual Classical
  and Einstein-Podolsky-Rosen Channels}},\ }\href
  {https://doi.org/10.1103/PhysRevLett.70.1895} {\bibfield  {journal} {\bibinfo
   {journal} {Phys. Rev. Lett.}\ }\textbf {\bibinfo {volume} {70}},\ \bibinfo
  {pages} {1895} (\bibinfo {year} {1993})}\BibitemShut {NoStop}%
\bibitem [{\citenamefont {Ekert}(1991)}]{Ekert1991}%
  \BibitemOpen
  \bibfield  {author} {\bibinfo {author} {\bibfnamefont {A.~K.}\ \bibnamefont
  {Ekert}},\ }\bibfield  {title} {\bibinfo {title} {{Quantum Cryptography Based
  on Bell's Theorem}},\ }\href {https://doi.org/10.1103/PhysRevLett.67.661}
  {\bibfield  {journal} {\bibinfo  {journal} {Phys. Rev. Lett.}\ }\textbf
  {\bibinfo {volume} {67}},\ \bibinfo {pages} {661} (\bibinfo {year}
  {1991})}\BibitemShut {NoStop}%
\bibitem [{\citenamefont {Briegel}\ \emph {et~al.}(2009)\citenamefont
  {Briegel}, \citenamefont {Browne}, \citenamefont {D{\"u}r}, \citenamefont
  {Raussendorf},\ and\ \citenamefont {den Nest}}]{Briegel2009}%
  \BibitemOpen
  \bibfield  {author} {\bibinfo {author} {\bibfnamefont {H.~J.}\ \bibnamefont
  {Briegel}}, \bibinfo {author} {\bibfnamefont {D.~E.}\ \bibnamefont {Browne}},
  \bibinfo {author} {\bibfnamefont {W.}~\bibnamefont {D{\"u}r}}, \bibinfo
  {author} {\bibfnamefont {R.}~\bibnamefont {Raussendorf}},\ and\ \bibinfo
  {author} {\bibfnamefont {M.~V.}\ \bibnamefont {den Nest}},\ }\bibfield
  {title} {\bibinfo {title} {{Measurement-Based Quantum Computation}},\ }\href
  {https://doi.org/10.1038/nphys1157} {\bibfield  {journal} {\bibinfo
  {journal} {Nat. Phys.}\ }\textbf {\bibinfo {volume} {5}},\ \bibinfo {pages}
  {19} (\bibinfo {year} {2009})}\BibitemShut {NoStop}%
\bibitem [{\citenamefont {Giovannetti}\ \emph {et~al.}(2004)\citenamefont
  {Giovannetti}, \citenamefont {Lloyd},\ and\ \citenamefont
  {Maccone}}]{Giovannetti2004}%
  \BibitemOpen
  \bibfield  {author} {\bibinfo {author} {\bibfnamefont {V.}~\bibnamefont
  {Giovannetti}}, \bibinfo {author} {\bibfnamefont {S.}~\bibnamefont {Lloyd}},\
  and\ \bibinfo {author} {\bibfnamefont {L.}~\bibnamefont {Maccone}},\
  }\bibfield  {title} {\bibinfo {title} {{Quantum-Enhanced Measurements:
  Beating the Standard Quantum Limit}},\ }\href
  {https://doi.org/10.1126/science.1104149} {\bibfield  {journal} {\bibinfo
  {journal} {Science}\ }\textbf {\bibinfo {volume} {306}},\ \bibinfo {pages}
  {1330} (\bibinfo {year} {2004})}\BibitemShut {NoStop}%
\bibitem [{\citenamefont {Dowling}\ and\ \citenamefont
  {Milburn}(2003)}]{Dowling2003}%
  \BibitemOpen
  \bibfield  {author} {\bibinfo {author} {\bibfnamefont {J.~P.}\ \bibnamefont
  {Dowling}}\ and\ \bibinfo {author} {\bibfnamefont {G.~J.}\ \bibnamefont
  {Milburn}},\ }\bibfield  {title} {\bibinfo {title} {{Quantum technology: the
  second quantum revolution}},\ }\href {https://doi.org/10.1098/rsta.2003.1227}
  {\bibfield  {journal} {\bibinfo  {journal} {Philos. trans., Math. phys. eng.
  sci.}\ }\textbf {\bibinfo {volume} {361}},\ \bibinfo {pages} {1655–1674}
  (\bibinfo {year} {2003})}\BibitemShut {NoStop}%
\bibitem [{\citenamefont {Deutsch}(2020)}]{Deutsch2020}%
  \BibitemOpen
  \bibfield  {author} {\bibinfo {author} {\bibfnamefont {I.~H.}\ \bibnamefont
  {Deutsch}},\ }\bibfield  {title} {\bibinfo {title} {{Harnessing the Power of
  the Second Quantum Revolution}},\ }\href
  {https://doi.org/10.1103/PRXQuantum.1.020101} {\bibfield  {journal} {\bibinfo
   {journal} {PRX Quantum}\ }\textbf {\bibinfo {volume} {1}},\ \bibinfo {pages}
  {020101} (\bibinfo {year} {2020})}\BibitemShut {NoStop}%
\bibitem [{\citenamefont {Huang}\ \emph {et~al.}(2026)\citenamefont {Huang},
  \citenamefont {Choi}, \citenamefont {McClean},\ and\ \citenamefont
  {Preskill}}]{Huang2026}%
  \BibitemOpen
  \bibfield  {author} {\bibinfo {author} {\bibfnamefont {H.-Y.}\ \bibnamefont
  {Huang}}, \bibinfo {author} {\bibfnamefont {S.}~\bibnamefont {Choi}},
  \bibinfo {author} {\bibfnamefont {J.~R.}\ \bibnamefont {McClean}},\ and\
  \bibinfo {author} {\bibfnamefont {J.}~\bibnamefont {Preskill}},\ }\bibfield
  {title} {\bibinfo {title} {{Vast World of Quantum Advantage}},\ }\href
  {https://doi.org/10.1103/tn89-g1xz} {\bibfield  {journal} {\bibinfo
  {journal} {Phys. Rev. X}\ }\textbf {\bibinfo {volume} {16}},\ \bibinfo
  {pages} {030501} (\bibinfo {year} {2026})}\BibitemShut {NoStop}%
\bibitem [{\citenamefont {Bell}(1964)}]{Bell1964}%
  \BibitemOpen
  \bibfield  {author} {\bibinfo {author} {\bibfnamefont {J.~S.}\ \bibnamefont
  {Bell}},\ }\bibfield  {title} {\bibinfo {title} {{On the Einstein Podolsky
  Rosen Paradox}},\ }\href
  {https://doi.org/10.1103/PhysicsPhysiqueFizika.1.195} {\bibfield  {journal}
  {\bibinfo  {journal} {Physics}\ }\textbf {\bibinfo {volume} {1}},\ \bibinfo
  {pages} {195} (\bibinfo {year} {1964})}\BibitemShut {NoStop}%
\bibitem [{\citenamefont {Bell}(1966)}]{Bell1966}%
  \BibitemOpen
  \bibfield  {author} {\bibinfo {author} {\bibfnamefont {J.~S.}\ \bibnamefont
  {Bell}},\ }\bibfield  {title} {\bibinfo {title} {{On the Problem of Hidden
  Variables in Quantum Mechanics}},\ }\href
  {https://doi.org/10.1103/RevModPhys.38.447} {\bibfield  {journal} {\bibinfo
  {journal} {Rev. Mod. Phys.}\ }\textbf {\bibinfo {volume} {38}},\ \bibinfo
  {pages} {447} (\bibinfo {year} {1966})}\BibitemShut {NoStop}%
\bibitem [{\citenamefont {Mermin}(1993)}]{Mermin1993}%
  \BibitemOpen
  \bibfield  {author} {\bibinfo {author} {\bibfnamefont {N.~D.}\ \bibnamefont
  {Mermin}},\ }\bibfield  {title} {\bibinfo {title} {{Hidden Variables and the
  Two Theorems of John Bell}},\ }\href
  {https://doi.org/10.1103/RevModPhys.65.803} {\bibfield  {journal} {\bibinfo
  {journal} {Rev. Mod. Phys.}\ }\textbf {\bibinfo {volume} {65}},\ \bibinfo
  {pages} {803} (\bibinfo {year} {1993})}\BibitemShut {NoStop}%
\bibitem [{\citenamefont {Wiseman}\ \emph {et~al.}(2007)\citenamefont
  {Wiseman}, \citenamefont {Jones},\ and\ \citenamefont
  {Doherty}}]{Wiseman2007}%
  \BibitemOpen
  \bibfield  {author} {\bibinfo {author} {\bibfnamefont {H.~M.}\ \bibnamefont
  {Wiseman}}, \bibinfo {author} {\bibfnamefont {S.~J.}\ \bibnamefont {Jones}},\
  and\ \bibinfo {author} {\bibfnamefont {A.~C.}\ \bibnamefont {Doherty}},\
  }\bibfield  {title} {\bibinfo {title} {{Steering, Entanglement, Nonlocality,
  and the Einstein-Podolsky-Rosen Paradox}},\ }\href
  {https://doi.org/10.1103/PhysRevLett.98.140402} {\bibfield  {journal}
  {\bibinfo  {journal} {Phys. Rev. Lett.}\ }\textbf {\bibinfo {volume} {98}},\
  \bibinfo {pages} {140402} (\bibinfo {year} {2007})}\BibitemShut {NoStop}%
\bibitem [{\citenamefont {Horodecki}\ \emph {et~al.}(2009)\citenamefont
  {Horodecki}, \citenamefont {Horodecki}, \citenamefont {Horodecki},\ and\
  \citenamefont {Horodecki}}]{Horodecki2009}%
  \BibitemOpen
  \bibfield  {author} {\bibinfo {author} {\bibfnamefont {R.}~\bibnamefont
  {Horodecki}}, \bibinfo {author} {\bibfnamefont {P.}~\bibnamefont
  {Horodecki}}, \bibinfo {author} {\bibfnamefont {M.}~\bibnamefont
  {Horodecki}},\ and\ \bibinfo {author} {\bibfnamefont {K.}~\bibnamefont
  {Horodecki}},\ }\bibfield  {title} {\bibinfo {title} {{Quantum
  Entanglement}},\ }\href {https://doi.org/10.1103/RevModPhys.81.865}
  {\bibfield  {journal} {\bibinfo  {journal} {Rev. Mod. Phys.}\ }\textbf
  {\bibinfo {volume} {81}},\ \bibinfo {pages} {865} (\bibinfo {year}
  {2009})}\BibitemShut {NoStop}%
\bibitem [{\citenamefont {Brunner}\ \emph {et~al.}(2014)\citenamefont
  {Brunner}, \citenamefont {Cavalcanti}, \citenamefont {Pironio}, \citenamefont
  {Scarani},\ and\ \citenamefont {Wehner}}]{Brunner2014}%
  \BibitemOpen
  \bibfield  {author} {\bibinfo {author} {\bibfnamefont {N.}~\bibnamefont
  {Brunner}}, \bibinfo {author} {\bibfnamefont {D.}~\bibnamefont {Cavalcanti}},
  \bibinfo {author} {\bibfnamefont {S.}~\bibnamefont {Pironio}}, \bibinfo
  {author} {\bibfnamefont {V.}~\bibnamefont {Scarani}},\ and\ \bibinfo {author}
  {\bibfnamefont {S.}~\bibnamefont {Wehner}},\ }\bibfield  {title} {\bibinfo
  {title} {{Bell Nonlocality}},\ }\href
  {https://doi.org/10.1103/RevModPhys.86.419} {\bibfield  {journal} {\bibinfo
  {journal} {Rev. Mod. Phys.}\ }\textbf {\bibinfo {volume} {86}},\ \bibinfo
  {pages} {419} (\bibinfo {year} {2014})}\BibitemShut {NoStop}%
\bibitem [{\citenamefont {{R. Uola and A. C. S. Costa and H. C. Nguyen and O.
  G{\"u}hne}}(2020)}]{Uola2020}%
  \BibitemOpen
  \bibfield  {author} {\bibinfo {author} {\bibnamefont {{R. Uola and A. C. S.
  Costa and H. C. Nguyen and O. G{\"u}hne}}},\ }\bibfield  {title} {\bibinfo
  {title} {Quantum steering},\ }\href
  {https://doi.org/10.1103/RevModPhys.92.015001} {\bibfield  {journal}
  {\bibinfo  {journal} {Rev. Mod. Phys.}\ }\textbf {\bibinfo {volume} {92}},\
  \bibinfo {pages} {015001} (\bibinfo {year} {2020})}\BibitemShut {NoStop}%
\bibitem [{\citenamefont {Werner}(1989)}]{Werner1989}%
  \BibitemOpen
  \bibfield  {author} {\bibinfo {author} {\bibfnamefont {R.~F.}\ \bibnamefont
  {Werner}},\ }\bibfield  {title} {\bibinfo {title} {{Quantum states with
  Einstein-Podolsky-Rosen correlations admitting a hidden-variable model}},\
  }\href {https://doi.org/10.1103/PhysRevA.40.4277} {\bibfield  {journal}
  {\bibinfo  {journal} {Phys. Rev. A}\ }\textbf {\bibinfo {volume} {40}},\
  \bibinfo {pages} {4277} (\bibinfo {year} {1989})}\BibitemShut {NoStop}%
\bibitem [{\citenamefont {Barrett}(2002)}]{Barrett2002}%
  \BibitemOpen
  \bibfield  {author} {\bibinfo {author} {\bibfnamefont {J.}~\bibnamefont
  {Barrett}},\ }\bibfield  {title} {\bibinfo {title} {{Nonsequential
  positive-operator-valued measurements on entangled mixed states do not always
  violate a Bell inequality}},\ }\href
  {https://doi.org/10.1103/PhysRevA.65.042302} {\bibfield  {journal} {\bibinfo
  {journal} {Phys. Rev. A}\ }\textbf {\bibinfo {volume} {65}},\ \bibinfo
  {pages} {042302} (\bibinfo {year} {2002})}\BibitemShut {NoStop}%
\bibitem [{\citenamefont {Masanes}(2006)}]{Masanes2006}%
  \BibitemOpen
  \bibfield  {author} {\bibinfo {author} {\bibfnamefont {L.}~\bibnamefont
  {Masanes}},\ }\bibfield  {title} {\bibinfo {title} {{Asymptotic Violation of
  Bell Inequalities and Distillability}},\ }\href
  {https://doi.org/10.1103/PhysRevLett.97.050503} {\bibfield  {journal}
  {\bibinfo  {journal} {Phys. Rev. Lett.}\ }\textbf {\bibinfo {volume} {97}},\
  \bibinfo {pages} {050503} (\bibinfo {year} {2006})}\BibitemShut {NoStop}%
\bibitem [{\citenamefont {Peres}(1999)}]{Peres1999}%
  \BibitemOpen
  \bibfield  {author} {\bibinfo {author} {\bibfnamefont {A.}~\bibnamefont
  {Peres}},\ }\bibfield  {title} {\bibinfo {title} {{All the Bell
  Inequalities}},\ }\href {https://doi.org/10.1023/a:1018816310000} {\bibfield
  {journal} {\bibinfo  {journal} {Found. Phys.}\ }\textbf {\bibinfo {volume}
  {29}},\ \bibinfo {pages} {589–614} (\bibinfo {year} {1999})}\BibitemShut
  {NoStop}%
\bibitem [{\citenamefont {Pusey}(2013)}]{Pusey2013}%
  \BibitemOpen
  \bibfield  {author} {\bibinfo {author} {\bibfnamefont {M.~F.}\ \bibnamefont
  {Pusey}},\ }\bibfield  {title} {\bibinfo {title} {{Negativity and steering: A
  stronger Peres conjecture}},\ }\href
  {https://doi.org/10.1103/PhysRevA.88.032313} {\bibfield  {journal} {\bibinfo
  {journal} {Phys. Rev. A}\ }\textbf {\bibinfo {volume} {88}},\ \bibinfo
  {pages} {032313} (\bibinfo {year} {2013})}\BibitemShut {NoStop}%
\bibitem [{\citenamefont {Skrzypczyk}\ \emph {et~al.}(2014)\citenamefont
  {Skrzypczyk}, \citenamefont {Navascu\'es},\ and\ \citenamefont
  {Cavalcanti}}]{Skrzypczyk2014}%
  \BibitemOpen
  \bibfield  {author} {\bibinfo {author} {\bibfnamefont {P.}~\bibnamefont
  {Skrzypczyk}}, \bibinfo {author} {\bibfnamefont {M.}~\bibnamefont
  {Navascu\'es}},\ and\ \bibinfo {author} {\bibfnamefont {D.}~\bibnamefont
  {Cavalcanti}},\ }\bibfield  {title} {\bibinfo {title} {{Quantifying
  Einstein-Podolsky-Rosen Steering}},\ }\href
  {https://doi.org/10.1103/PhysRevLett.112.180404} {\bibfield  {journal}
  {\bibinfo  {journal} {Phys. Rev. Lett.}\ }\textbf {\bibinfo {volume} {112}},\
  \bibinfo {pages} {180404} (\bibinfo {year} {2014})}\BibitemShut {NoStop}%
\bibitem [{\citenamefont {D{\"u}r}(2001)}]{Dur2001}%
  \BibitemOpen
  \bibfield  {author} {\bibinfo {author} {\bibfnamefont {W.}~\bibnamefont
  {D{\"u}r}},\ }\bibfield  {title} {\bibinfo {title} {Multipartite bound
  entangled states that violate bell's inequality},\ }\href
  {https://doi.org/10.1103/PhysRevLett.87.230402} {\bibfield  {journal}
  {\bibinfo  {journal} {Phys. Rev. Lett.}\ }\textbf {\bibinfo {volume} {87}},\
  \bibinfo {pages} {230402} (\bibinfo {year} {2001})}\BibitemShut {NoStop}%
\bibitem [{\citenamefont {V{\'e}rtesi}\ and\ \citenamefont
  {Brunner}(2012)}]{Vertesi2012}%
  \BibitemOpen
  \bibfield  {author} {\bibinfo {author} {\bibfnamefont {T.}~\bibnamefont
  {V{\'e}rtesi}}\ and\ \bibinfo {author} {\bibfnamefont {N.}~\bibnamefont
  {Brunner}},\ }\bibfield  {title} {\bibinfo {title} {Quantum nonlocality does
  not imply entanglement distillability},\ }\href
  {https://doi.org/10.1103/PhysRevLett.108.030403} {\bibfield  {journal}
  {\bibinfo  {journal} {Phys. Rev. Lett.}\ }\textbf {\bibinfo {volume} {108}},\
  \bibinfo {pages} {030403} (\bibinfo {year} {2012})}\BibitemShut {NoStop}%
\bibitem [{\citenamefont {Moroder}\ \emph {et~al.}(2014)\citenamefont
  {Moroder}, \citenamefont {Gittsovich}, \citenamefont {Huber},\ and\
  \citenamefont {G\"uhne}}]{Moroder2014}%
  \BibitemOpen
  \bibfield  {author} {\bibinfo {author} {\bibfnamefont {T.}~\bibnamefont
  {Moroder}}, \bibinfo {author} {\bibfnamefont {O.}~\bibnamefont {Gittsovich}},
  \bibinfo {author} {\bibfnamefont {M.}~\bibnamefont {Huber}},\ and\ \bibinfo
  {author} {\bibfnamefont {O.}~\bibnamefont {G\"uhne}},\ }\bibfield  {title}
  {\bibinfo {title} {{Steering Bound Entangled States: A Counterexample to the
  Stronger Peres Conjecture}},\ }\href
  {https://doi.org/10.1103/PhysRevLett.113.050404} {\bibfield  {journal}
  {\bibinfo  {journal} {Phys. Rev. Lett.}\ }\textbf {\bibinfo {volume} {113}},\
  \bibinfo {pages} {050404} (\bibinfo {year} {2014})}\BibitemShut {NoStop}%
\bibitem [{\citenamefont {Vértesi}\ and\ \citenamefont
  {Brunner}(2014)}]{Vertesi2014}%
  \BibitemOpen
  \bibfield  {author} {\bibinfo {author} {\bibfnamefont {T.}~\bibnamefont
  {Vértesi}}\ and\ \bibinfo {author} {\bibfnamefont {N.}~\bibnamefont
  {Brunner}},\ }\bibfield  {title} {\bibinfo {title} {{Disproving the Peres
  conjecture by showing Bell nonlocality from bound entanglement}},\ }\href
  {https://doi.org/10.1038/ncomms6297} {\bibfield  {journal} {\bibinfo
  {journal} {Nat. Commun.}\ }\textbf {\bibinfo {volume} {5}},\ \bibinfo {pages}
  {5297} (\bibinfo {year} {2014})}\BibitemShut {NoStop}%
\bibitem [{\citenamefont {Chen}\ and\ \citenamefont
  {{\DJ}okovi{\'c}}(2011{\natexlab{a}})}]{Chen2011a}%
  \BibitemOpen
  \bibfield  {author} {\bibinfo {author} {\bibfnamefont {L.}~\bibnamefont
  {Chen}}\ and\ \bibinfo {author} {\bibfnamefont {D.~{\v{Z}}.}\ \bibnamefont
  {{\DJ}okovi{\'c}}},\ }\bibfield  {title} {\bibinfo {title} {Distillability
  and ppt entanglement of low-rank quantum states},\ }\href
  {https://doi.org/10.1088/1751-8113/44/28/285303} {\bibfield  {journal}
  {\bibinfo  {journal} {J. Phys. A: Math. Theory}\ }\textbf {\bibinfo {volume}
  {44}},\ \bibinfo {pages} {285303} (\bibinfo {year}
  {2011}{\natexlab{a}})}\BibitemShut {NoStop}%
\bibitem [{\citenamefont {Chen}\ and\ \citenamefont
  {{\DJ}okovi{\'c}}(2011{\natexlab{b}})}]{Chen2011b}%
  \BibitemOpen
  \bibfield  {author} {\bibinfo {author} {\bibfnamefont {L.}~\bibnamefont
  {Chen}}\ and\ \bibinfo {author} {\bibfnamefont {D.~{\v{Z}}.}\ \bibnamefont
  {{\DJ}okovi{\'c}}},\ }\bibfield  {title} {\bibinfo {title} {Description of
  rank four entangled states of two qutrits having positive partial
  transpose},\ }\href {https://doi.org/10.1063/1.3663837} {\bibfield  {journal}
  {\bibinfo  {journal} {J. Math. Phys.}\ }\textbf {\bibinfo {volume} {52}},\
  \bibinfo {pages} {122203} (\bibinfo {year} {2011}{\natexlab{b}})}\BibitemShut
  {NoStop}%
\bibitem [{\citenamefont {Chen}\ and\ \citenamefont
  {{\DJ}okovi{\'c}}(2012)}]{ChenDjokovic2012classes}%
  \BibitemOpen
  \bibfield  {author} {\bibinfo {author} {\bibfnamefont {L.}~\bibnamefont
  {Chen}}\ and\ \bibinfo {author} {\bibfnamefont {D.~Z.}\ \bibnamefont
  {{\DJ}okovi{\'c}}},\ }\bibfield  {title} {\bibinfo {title} {Equivalence
  classes and canonical forms for two-qutrit entangled states of rank four
  having positive partial transpose},\ }\href
  {https://doi.org/10.1063/1.4757276} {\bibfield  {journal} {\bibinfo
  {journal} {J. Math. Phys.}\ }\textbf {\bibinfo {volume} {53}},\ \bibinfo
  {pages} {102205} (\bibinfo {year} {2012})}\BibitemShut {NoStop}%
\bibitem [{\citenamefont {Chen}\ and\ \citenamefont
  {{\DJ}okovi{\'c}}(2013)}]{Chen2013b}%
  \BibitemOpen
  \bibfield  {author} {\bibinfo {author} {\bibfnamefont {L.}~\bibnamefont
  {Chen}}\ and\ \bibinfo {author} {\bibfnamefont {D.~{\v{Z}}.}\ \bibnamefont
  {{\DJ}okovi{\'c}}},\ }\bibfield  {title} {\bibinfo {title} {Separability
  problem for multipartite states of rank at most 4},\ }\href
  {https://doi.org/10.1088/1751-8113/46/27/275304} {\bibfield  {journal}
  {\bibinfo  {journal} {J. Phys. A: Math. Theory}\ }\textbf {\bibinfo {volume}
  {46}},\ \bibinfo {pages} {275304} (\bibinfo {year} {2013})}\BibitemShut
  {NoStop}%
\bibitem [{\citenamefont {krishnan S~V}\ \emph {et~al.}(2021)\citenamefont
  {krishnan S~V}, \citenamefont {Ranjan},\ and\ \citenamefont
  {Banik}}]{KrishnanRanjanBanik2021}%
  \BibitemOpen
  \bibfield  {author} {\bibinfo {author} {\bibfnamefont {H.}~\bibnamefont
  {krishnan S~V}}, \bibinfo {author} {\bibfnamefont {A.}~\bibnamefont
  {Ranjan}},\ and\ \bibinfo {author} {\bibfnamefont {M.}~\bibnamefont
  {Banik}},\ }\bibfield  {title} {\bibinfo {title} {State space structure of
  tripartite quantum systems},\ }\href
  {https://doi.org/10.1103/PhysRevA.104.022437} {\bibfield  {journal} {\bibinfo
   {journal} {Phys. Rev. A}\ }\textbf {\bibinfo {volume} {104}},\ \bibinfo
  {pages} {022437} (\bibinfo {year} {2021})}\BibitemShut {NoStop}%
\bibitem [{\citenamefont {G\"{u}hne}\ and\ \citenamefont
  {Tóth}(2009)}]{Ghune2009}%
  \BibitemOpen
  \bibfield  {author} {\bibinfo {author} {\bibfnamefont {O.}~\bibnamefont
  {G\"{u}hne}}\ and\ \bibinfo {author} {\bibfnamefont {G.}~\bibnamefont
  {Tóth}},\ }\bibfield  {title} {\bibinfo {title} {Entanglement detection},\
  }\href {https://doi.org/10.1016/j.physrep.2009.02.004} {\bibfield  {journal}
  {\bibinfo  {journal} {Phys. Rep.}\ }\textbf {\bibinfo {volume} {474}},\
  \bibinfo {pages} {1–75} (\bibinfo {year} {2009})}\BibitemShut {NoStop}%
\bibitem [{\citenamefont {Jungnitsch}\ \emph {et~al.}(2011)\citenamefont
  {Jungnitsch}, \citenamefont {Moroder},\ and\ \citenamefont
  {G\"uhne}}]{Jungnitsch2011}%
  \BibitemOpen
  \bibfield  {author} {\bibinfo {author} {\bibfnamefont {B.}~\bibnamefont
  {Jungnitsch}}, \bibinfo {author} {\bibfnamefont {T.}~\bibnamefont
  {Moroder}},\ and\ \bibinfo {author} {\bibfnamefont {O.}~\bibnamefont
  {G\"uhne}},\ }\bibfield  {title} {\bibinfo {title} {{Taming Multiparticle
  Entanglement}},\ }\href {https://doi.org/10.1103/PhysRevLett.106.190502}
  {\bibfield  {journal} {\bibinfo  {journal} {Phys. Rev. Lett.}\ }\textbf
  {\bibinfo {volume} {106}},\ \bibinfo {pages} {190502} (\bibinfo {year}
  {2011})}\BibitemShut {NoStop}%
\bibitem [{\citenamefont {Bennett}\ \emph
  {et~al.}(1999{\natexlab{a}})\citenamefont {Bennett}, \citenamefont
  {DiVincenzo}, \citenamefont {Mor}, \citenamefont {Shor}, \citenamefont
  {Smolin},\ and\ \citenamefont {Terhal}}]{Bennett1999}%
  \BibitemOpen
  \bibfield  {author} {\bibinfo {author} {\bibfnamefont {C.~H.}\ \bibnamefont
  {Bennett}}, \bibinfo {author} {\bibfnamefont {D.~P.}\ \bibnamefont
  {DiVincenzo}}, \bibinfo {author} {\bibfnamefont {T.}~\bibnamefont {Mor}},
  \bibinfo {author} {\bibfnamefont {P.~W.}\ \bibnamefont {Shor}}, \bibinfo
  {author} {\bibfnamefont {J.~A.}\ \bibnamefont {Smolin}},\ and\ \bibinfo
  {author} {\bibfnamefont {B.~M.}\ \bibnamefont {Terhal}},\ }\bibfield  {title}
  {\bibinfo {title} {{Unextendible Product Bases and Bound Entanglement}},\
  }\href {https://doi.org/10.1103/PhysRevLett.82.5385} {\bibfield  {journal}
  {\bibinfo  {journal} {Phys. Rev. Lett.}\ }\textbf {\bibinfo {volume} {82}},\
  \bibinfo {pages} {5385} (\bibinfo {year} {1999}{\natexlab{a}})}\BibitemShut
  {NoStop}%
\bibitem [{\citenamefont {Ha}\ and\ \citenamefont
  {Kye}(2016{\natexlab{a}})}]{Ha2016}%
  \BibitemOpen
  \bibfield  {author} {\bibinfo {author} {\bibfnamefont {K.-C.}\ \bibnamefont
  {Ha}}\ and\ \bibinfo {author} {\bibfnamefont {S.-H.}\ \bibnamefont {Kye}},\
  }\bibfield  {title} {\bibinfo {title} {{Construction of three-qubit genuine
  entanglement with bipartite positive partial transposes}},\ }\href
  {https://doi.org/10.1103/PhysRevA.93.032315} {\bibfield  {journal} {\bibinfo
  {journal} {Phys. Rev. A}\ }\textbf {\bibinfo {volume} {93}},\ \bibinfo
  {pages} {032315} (\bibinfo {year} {2016}{\natexlab{a}})}\BibitemShut
  {NoStop}%
\bibitem [{\citenamefont {Novo}\ \emph {et~al.}(2013)\citenamefont {Novo},
  \citenamefont {Moroder},\ and\ \citenamefont {G\"uhne}}]{Novo2013}%
  \BibitemOpen
  \bibfield  {author} {\bibinfo {author} {\bibfnamefont {L.}~\bibnamefont
  {Novo}}, \bibinfo {author} {\bibfnamefont {T.}~\bibnamefont {Moroder}},\ and\
  \bibinfo {author} {\bibfnamefont {O.}~\bibnamefont {G\"uhne}},\ }\bibfield
  {title} {\bibinfo {title} {{Genuine multiparticle entanglement of
  permutationally invariant states}},\ }\href
  {https://doi.org/10.1103/PhysRevA.88.012305} {\bibfield  {journal} {\bibinfo
  {journal} {Phys. Rev. A}\ }\textbf {\bibinfo {volume} {88}},\ \bibinfo
  {pages} {012305} (\bibinfo {year} {2013})}\BibitemShut {NoStop}%
\bibitem [{\citenamefont {Svetlichny}(1987)}]{Svetlichny1987}%
  \BibitemOpen
  \bibfield  {author} {\bibinfo {author} {\bibfnamefont {G.}~\bibnamefont
  {Svetlichny}},\ }\bibfield  {title} {\bibinfo {title} {{Distinguishing
  Three-Body from Two-Body Nonseparability by a Bell-Type Inequality}},\ }\href
  {https://doi.org/10.1103/PhysRevD.35.3066} {\bibfield  {journal} {\bibinfo
  {journal} {Phys. Rev. D}\ }\textbf {\bibinfo {volume} {35}},\ \bibinfo
  {pages} {3066} (\bibinfo {year} {1987})}\BibitemShut {NoStop}%
\bibitem [{\citenamefont {Gallego}\ \emph {et~al.}(2012)\citenamefont
  {Gallego}, \citenamefont {W\"urflinger}, \citenamefont {Ac\'{\i}n},\ and\
  \citenamefont {Navascu\'es}}]{Gallego2012}%
  \BibitemOpen
  \bibfield  {author} {\bibinfo {author} {\bibfnamefont {R.}~\bibnamefont
  {Gallego}}, \bibinfo {author} {\bibfnamefont {L.~E.}\ \bibnamefont
  {W\"urflinger}}, \bibinfo {author} {\bibfnamefont {A.}~\bibnamefont
  {Ac\'{\i}n}},\ and\ \bibinfo {author} {\bibfnamefont {M.}~\bibnamefont
  {Navascu\'es}},\ }\bibfield  {title} {\bibinfo {title} {{Operational
  Framework for Nonlocality}},\ }\href
  {https://doi.org/10.1103/PhysRevLett.109.070401} {\bibfield  {journal}
  {\bibinfo  {journal} {Phys. Rev. Lett.}\ }\textbf {\bibinfo {volume} {109}},\
  \bibinfo {pages} {070401} (\bibinfo {year} {2012})}\BibitemShut {NoStop}%
\bibitem [{\citenamefont {Bancal}\ \emph {et~al.}(2013)\citenamefont {Bancal},
  \citenamefont {Barrett}, \citenamefont {Gisin},\ and\ \citenamefont
  {Pironio}}]{Bancal2013}%
  \BibitemOpen
  \bibfield  {author} {\bibinfo {author} {\bibfnamefont {J.-D.}\ \bibnamefont
  {Bancal}}, \bibinfo {author} {\bibfnamefont {J.}~\bibnamefont {Barrett}},
  \bibinfo {author} {\bibfnamefont {N.}~\bibnamefont {Gisin}},\ and\ \bibinfo
  {author} {\bibfnamefont {S.}~\bibnamefont {Pironio}},\ }\bibfield  {title}
  {\bibinfo {title} {{Definitions of multipartite nonlocality}},\ }\href
  {https://doi.org/10.1103/PhysRevA.88.014102} {\bibfield  {journal} {\bibinfo
  {journal} {Phys. Rev. A}\ }\textbf {\bibinfo {volume} {88}},\ \bibinfo
  {pages} {014102} (\bibinfo {year} {2013})}\BibitemShut {NoStop}%
\bibitem [{\citenamefont {Dutta}\ \emph {et~al.}(2020)\citenamefont {Dutta},
  \citenamefont {Mukherjee},\ and\ \citenamefont {Banik}}]{Dutta2020}%
  \BibitemOpen
  \bibfield  {author} {\bibinfo {author} {\bibfnamefont {S.}~\bibnamefont
  {Dutta}}, \bibinfo {author} {\bibfnamefont {A.}~\bibnamefont {Mukherjee}},\
  and\ \bibinfo {author} {\bibfnamefont {M.}~\bibnamefont {Banik}},\ }\bibfield
   {title} {\bibinfo {title} {{Operational characterization of multipartite
  nonlocal correlations}},\ }\href
  {https://doi.org/10.1103/PhysRevA.102.052218} {\bibfield  {journal} {\bibinfo
   {journal} {Phys. Rev. A}\ }\textbf {\bibinfo {volume} {102}},\ \bibinfo
  {pages} {052218} (\bibinfo {year} {2020})}\BibitemShut {NoStop}%
\bibitem [{\citenamefont {DiVincenzo}\ \emph {et~al.}(2003)\citenamefont
  {DiVincenzo}, \citenamefont {Mor}, \citenamefont {Shor}, \citenamefont
  {Smolin},\ and\ \citenamefont {Terhal}}]{DiVincenzo2003}%
  \BibitemOpen
  \bibfield  {author} {\bibinfo {author} {\bibfnamefont {D.~P.}\ \bibnamefont
  {DiVincenzo}}, \bibinfo {author} {\bibfnamefont {T.}~\bibnamefont {Mor}},
  \bibinfo {author} {\bibfnamefont {P.~W.}\ \bibnamefont {Shor}}, \bibinfo
  {author} {\bibfnamefont {J.~A.}\ \bibnamefont {Smolin}},\ and\ \bibinfo
  {author} {\bibfnamefont {B.~M.}\ \bibnamefont {Terhal}},\ }\bibfield  {title}
  {\bibinfo {title} {{Unextendible Product Bases, Uncompletable Product Bases
  and Bound Entanglement}},\ }\href {https://doi.org/10.1007/s00220-003-0877-6}
  {\bibfield  {journal} {\bibinfo  {journal} {Commun. Math. Phys.}\ }\textbf
  {\bibinfo {volume} {238}},\ \bibinfo {pages} {379} (\bibinfo {year}
  {2003})}\BibitemShut {NoStop}%
\bibitem [{\citenamefont {Parthasarathy}(2004)}]{Parthasarathy2004}%
  \BibitemOpen
  \bibfield  {author} {\bibinfo {author} {\bibfnamefont {K.~R.}\ \bibnamefont
  {Parthasarathy}},\ }\bibfield  {title} {\bibinfo {title} {{On the Maximal
  Dimension of a Completely Entangled Subspace for Finite Level Quantum
  Systems}},\ }\href {https://doi.org/10.1007/BF02829441} {\bibfield  {journal}
  {\bibinfo  {journal} {Proc. Indian Acad. Sci.: Math. Sci.}\ }\textbf
  {\bibinfo {volume} {114}},\ \bibinfo {pages} {365} (\bibinfo {year}
  {2004})}\BibitemShut {NoStop}%
\bibitem [{\citenamefont {Bennett}\ \emph
  {et~al.}(1999{\natexlab{b}})\citenamefont {Bennett}, \citenamefont
  {DiVincenzo}, \citenamefont {Fuchs}, \citenamefont {Mor}, \citenamefont
  {Rains}, \citenamefont {Shor}, \citenamefont {Smolin},\ and\ \citenamefont
  {Wootters}}]{Bennett1999NWE}%
  \BibitemOpen
  \bibfield  {author} {\bibinfo {author} {\bibfnamefont {C.~H.}\ \bibnamefont
  {Bennett}}, \bibinfo {author} {\bibfnamefont {D.~P.}\ \bibnamefont
  {DiVincenzo}}, \bibinfo {author} {\bibfnamefont {C.~A.}\ \bibnamefont
  {Fuchs}}, \bibinfo {author} {\bibfnamefont {T.}~\bibnamefont {Mor}}, \bibinfo
  {author} {\bibfnamefont {E.}~\bibnamefont {Rains}}, \bibinfo {author}
  {\bibfnamefont {P.~W.}\ \bibnamefont {Shor}}, \bibinfo {author}
  {\bibfnamefont {J.~A.}\ \bibnamefont {Smolin}},\ and\ \bibinfo {author}
  {\bibfnamefont {W.~K.}\ \bibnamefont {Wootters}},\ }\bibfield  {title}
  {\bibinfo {title} {{Quantum Nonlocality without Entanglement}},\ }\href
  {https://doi.org/10.1103/PhysRevA.59.1070} {\bibfield  {journal} {\bibinfo
  {journal} {Phys. Rev. A}\ }\textbf {\bibinfo {volume} {59}},\ \bibinfo
  {pages} {1070} (\bibinfo {year} {1999}{\natexlab{b}})}\BibitemShut {NoStop}%
\bibitem [{\citenamefont {Walgate}\ and\ \citenamefont
  {Hardy}(2002)}]{Walgate2002}%
  \BibitemOpen
  \bibfield  {author} {\bibinfo {author} {\bibfnamefont {J.}~\bibnamefont
  {Walgate}}\ and\ \bibinfo {author} {\bibfnamefont {L.}~\bibnamefont
  {Hardy}},\ }\bibfield  {title} {\bibinfo {title} {{Nonlocality, Asymmetry,
  and Distinguishing Bipartite States}},\ }\href
  {https://doi.org/10.1103/PhysRevLett.89.147901} {\bibfield  {journal}
  {\bibinfo  {journal} {Phys. Rev. Lett.}\ }\textbf {\bibinfo {volume} {89}},\
  \bibinfo {pages} {147901} (\bibinfo {year} {2002})}\BibitemShut {NoStop}%
\bibitem [{\citenamefont {Halder}\ \emph {et~al.}(2019)\citenamefont {Halder},
  \citenamefont {Banik}, \citenamefont {Agrawal},\ and\ \citenamefont
  {Bandyopadhyay}}]{Halder2019}%
  \BibitemOpen
  \bibfield  {author} {\bibinfo {author} {\bibfnamefont {S.}~\bibnamefont
  {Halder}}, \bibinfo {author} {\bibfnamefont {M.}~\bibnamefont {Banik}},
  \bibinfo {author} {\bibfnamefont {S.}~\bibnamefont {Agrawal}},\ and\ \bibinfo
  {author} {\bibfnamefont {S.}~\bibnamefont {Bandyopadhyay}},\ }\bibfield
  {title} {\bibinfo {title} {{Strong Quantum Nonlocality without
  Entanglement}},\ }\href {https://doi.org/10.1103/PhysRevLett.122.040403}
  {\bibfield  {journal} {\bibinfo  {journal} {Phys. Rev. Lett.}\ }\textbf
  {\bibinfo {volume} {122}},\ \bibinfo {pages} {040403} (\bibinfo {year}
  {2019})}\BibitemShut {NoStop}%
\bibitem [{\citenamefont {Bhattacharya}\ \emph {et~al.}(2020)\citenamefont
  {Bhattacharya}, \citenamefont {Saha}, \citenamefont {Guha},\ and\
  \citenamefont {Banik}}]{Bhattacharya2020NWE}%
  \BibitemOpen
  \bibfield  {author} {\bibinfo {author} {\bibfnamefont {S.~S.}\ \bibnamefont
  {Bhattacharya}}, \bibinfo {author} {\bibfnamefont {S.}~\bibnamefont {Saha}},
  \bibinfo {author} {\bibfnamefont {T.}~\bibnamefont {Guha}},\ and\ \bibinfo
  {author} {\bibfnamefont {M.}~\bibnamefont {Banik}},\ }\bibfield  {title}
  {\bibinfo {title} {{Nonlocality without Entanglement: Quantum Theory and
  Beyond}},\ }\href {https://doi.org/10.1103/PhysRevResearch.2.012068}
  {\bibfield  {journal} {\bibinfo  {journal} {Phys. Rev. Research}\ }\textbf
  {\bibinfo {volume} {2}},\ \bibinfo {pages} {012068} (\bibinfo {year}
  {2020})}\BibitemShut {NoStop}%
\bibitem [{\citenamefont {Agrawal}\ \emph {et~al.}(2019)\citenamefont
  {Agrawal}, \citenamefont {Halder},\ and\ \citenamefont
  {Banik}}]{Agrawal2019}%
  \BibitemOpen
  \bibfield  {author} {\bibinfo {author} {\bibfnamefont {S.}~\bibnamefont
  {Agrawal}}, \bibinfo {author} {\bibfnamefont {S.}~\bibnamefont {Halder}},\
  and\ \bibinfo {author} {\bibfnamefont {M.}~\bibnamefont {Banik}},\ }\bibfield
   {title} {\bibinfo {title} {{Genuinely entangled subspace with
  all-encompassing distillable entanglement across every bipartition}},\ }\href
  {https://doi.org/10.1103/PhysRevA.99.032335} {\bibfield  {journal} {\bibinfo
  {journal} {Phys. Rev. A}\ }\textbf {\bibinfo {volume} {99}},\ \bibinfo
  {pages} {032335} (\bibinfo {year} {2019})}\BibitemShut {NoStop}%
\bibitem [{\citenamefont {Rout}\ \emph {et~al.}(2019)\citenamefont {Rout},
  \citenamefont {Maity}, \citenamefont {Mukherjee}, \citenamefont {Halder},\
  and\ \citenamefont {Banik}}]{Rout2019}%
  \BibitemOpen
  \bibfield  {author} {\bibinfo {author} {\bibfnamefont {S.}~\bibnamefont
  {Rout}}, \bibinfo {author} {\bibfnamefont {A.~G.}\ \bibnamefont {Maity}},
  \bibinfo {author} {\bibfnamefont {A.}~\bibnamefont {Mukherjee}}, \bibinfo
  {author} {\bibfnamefont {S.}~\bibnamefont {Halder}},\ and\ \bibinfo {author}
  {\bibfnamefont {M.}~\bibnamefont {Banik}},\ }\bibfield  {title} {\bibinfo
  {title} {{Genuinely nonlocal product bases: Classification and
  entanglement-assisted discrimination}},\ }\href
  {https://doi.org/10.1103/PhysRevA.100.032321} {\bibfield  {journal} {\bibinfo
   {journal} {Phys. Rev. A}\ }\textbf {\bibinfo {volume} {100}},\ \bibinfo
  {pages} {032321} (\bibinfo {year} {2019})}\BibitemShut {NoStop}%
\bibitem [{\citenamefont {{\'S}liwa}(2003)}]{Sliwa2003}%
  \BibitemOpen
  \bibfield  {author} {\bibinfo {author} {\bibfnamefont {C.}~\bibnamefont
  {{\'S}liwa}},\ }\bibfield  {title} {\bibinfo {title} {Symmetries of the bell
  correlation inequalities},\ }\href
  {https://doi.org/10.1016/S0375-9601(03)01115-0} {\bibfield  {journal}
  {\bibinfo  {journal} {Physics Letters A}\ }\textbf {\bibinfo {volume}
  {317}},\ \bibinfo {pages} {165} (\bibinfo {year} {2003})}\BibitemShut
  {NoStop}%
\bibitem [{\citenamefont {Epping}\ and\ \citenamefont
  {Brukner}(2013)}]{Epping2013}%
  \BibitemOpen
  \bibfield  {author} {\bibinfo {author} {\bibfnamefont {M.}~\bibnamefont
  {Epping}}\ and\ \bibinfo {author} {\bibfnamefont {i.~c.~v.}\ \bibnamefont
  {Brukner}},\ }\bibfield  {title} {\bibinfo {title} {Bound entanglement helps
  to reduce communication complexity},\ }\href
  {https://doi.org/10.1103/PhysRevA.87.032305} {\bibfield  {journal} {\bibinfo
  {journal} {Phys. Rev. A}\ }\textbf {\bibinfo {volume} {87}},\ \bibinfo
  {pages} {032305} (\bibinfo {year} {2013})}\BibitemShut {NoStop}%
\bibitem [{\citenamefont {Kraus}\ \emph {et~al.}(2000)\citenamefont {Kraus},
  \citenamefont {Cirac}, \citenamefont {Karnas},\ and\ \citenamefont
  {Lewenstein}}]{Kraus2000}%
  \BibitemOpen
  \bibfield  {author} {\bibinfo {author} {\bibfnamefont {B.}~\bibnamefont
  {Kraus}}, \bibinfo {author} {\bibfnamefont {J.~I.}\ \bibnamefont {Cirac}},
  \bibinfo {author} {\bibfnamefont {S.}~\bibnamefont {Karnas}},\ and\ \bibinfo
  {author} {\bibfnamefont {M.}~\bibnamefont {Lewenstein}},\ }\bibfield  {title}
  {\bibinfo {title} {Separability in $2\times n$ composite quantum systems},\
  }\href {https://doi.org/10.1103/PhysRevA.61.062302} {\bibfield  {journal}
  {\bibinfo  {journal} {Phys. Rev. A}\ }\textbf {\bibinfo {volume} {61}},\
  \bibinfo {pages} {062302} (\bibinfo {year} {2000})}\BibitemShut {NoStop}%
\bibitem [{\citenamefont {Cheng}\ and\ \citenamefont
  {Chen}(2026)}]{ChengChen2026}%
  \BibitemOpen
  \bibfield  {author} {\bibinfo {author} {\bibfnamefont {Y.}~\bibnamefont
  {Cheng}}\ and\ \bibinfo {author} {\bibfnamefont {L.}~\bibnamefont {Chen}},\
  }\href@noop {} {\bibinfo {title} {Construction of three-qubit
  positive-partial-transpose entangled states of rank four}} (\bibinfo {year}
  {2026}),\ \Eprint {https://arxiv.org/abs/2605.19530} {arXiv:2605.19530
  [quant-ph]} \BibitemShut {NoStop}%
\bibitem [{\citenamefont {Bravyi}(2004)}]{Bravyi2004}%
  \BibitemOpen
  \bibfield  {author} {\bibinfo {author} {\bibfnamefont {S.}~\bibnamefont
  {Bravyi}},\ }\bibfield  {title} {\bibinfo {title} {Unextendible product bases
  and locally unconvertible bound entangled states},\ }\href
  {https://doi.org/10.1007/s11128-004-7076-z} {\bibfield  {journal} {\bibinfo
  {journal} {Quantum Information Processing}\ }\textbf {\bibinfo {volume}
  {3}},\ \bibinfo {pages} {309} (\bibinfo {year} {2004})}\BibitemShut {NoStop}%
\bibitem [{\citenamefont {Tendick}\ \emph {et~al.}(2020)\citenamefont
  {Tendick}, \citenamefont {Kampermann},\ and\ \citenamefont
  {Bru{\ss}}}]{Tendick2020}%
  \BibitemOpen
  \bibfield  {author} {\bibinfo {author} {\bibfnamefont {L.}~\bibnamefont
  {Tendick}}, \bibinfo {author} {\bibfnamefont {H.}~\bibnamefont
  {Kampermann}},\ and\ \bibinfo {author} {\bibfnamefont {D.}~\bibnamefont
  {Bru{\ss}}},\ }\bibfield  {title} {\bibinfo {title} {Activation of
  nonlocality in bound entanglement},\ }\href
  {https://doi.org/10.1103/PhysRevLett.124.050401} {\bibfield  {journal}
  {\bibinfo  {journal} {Phys. Rev. Lett.}\ }\textbf {\bibinfo {volume} {124}},\
  \bibinfo {pages} {050401} (\bibinfo {year} {2020})}\BibitemShut {NoStop}%
\bibitem [{\citenamefont {Terhal}\ \emph {et~al.}(2003)\citenamefont {Terhal},
  \citenamefont {Doherty},\ and\ \citenamefont {Schwab}}]{Terhal2003}%
  \BibitemOpen
  \bibfield  {author} {\bibinfo {author} {\bibfnamefont {B.~M.}\ \bibnamefont
  {Terhal}}, \bibinfo {author} {\bibfnamefont {A.~C.}\ \bibnamefont
  {Doherty}},\ and\ \bibinfo {author} {\bibfnamefont {D.}~\bibnamefont
  {Schwab}},\ }\bibfield  {title} {\bibinfo {title} {Symmetric extensions of
  quantum states and local hidden variable theories},\ }\href
  {https://doi.org/10.1103/PhysRevLett.90.157903} {\bibfield  {journal}
  {\bibinfo  {journal} {Phys. Rev. Lett.}\ }\textbf {\bibinfo {volume} {90}},\
  \bibinfo {pages} {157903} (\bibinfo {year} {2003})}\BibitemShut {NoStop}%
\bibitem [{\citenamefont {Steensgaard~Garberg}\ \emph
  {et~al.}(2013)\citenamefont {Steensgaard~Garberg}, \citenamefont {Irgens},\
  and\ \citenamefont {Myrheim}}]{Garberg2013}%
  \BibitemOpen
  \bibfield  {author} {\bibinfo {author} {\bibfnamefont {O.}~\bibnamefont
  {Steensgaard~Garberg}}, \bibinfo {author} {\bibfnamefont {B.}~\bibnamefont
  {Irgens}},\ and\ \bibinfo {author} {\bibfnamefont {J.}~\bibnamefont
  {Myrheim}},\ }\bibfield  {title} {\bibinfo {title} {Extremal states of
  positive partial transpose in a system of three qubits},\ }\href
  {https://doi.org/10.1103/PhysRevA.87.032302} {\bibfield  {journal} {\bibinfo
  {journal} {Phys. Rev. A}\ }\textbf {\bibinfo {volume} {87}},\ \bibinfo
  {pages} {032302} (\bibinfo {year} {2013})}\BibitemShut {NoStop}%
\bibitem [{\citenamefont {Ha}\ and\ \citenamefont
  {Kye}(2016{\natexlab{b}})}]{HaKye2016}%
  \BibitemOpen
  \bibfield  {author} {\bibinfo {author} {\bibfnamefont {K.-C.}\ \bibnamefont
  {Ha}}\ and\ \bibinfo {author} {\bibfnamefont {S.-H.}\ \bibnamefont {Kye}},\
  }\bibfield  {title} {\bibinfo {title} {Construction of three-qubit genuine
  entanglement with bipartite positive partial transposes},\ }\href
  {https://doi.org/10.1103/PhysRevA.93.032315} {\bibfield  {journal} {\bibinfo
  {journal} {Phys. Rev. A}\ }\textbf {\bibinfo {volume} {93}},\ \bibinfo
  {pages} {032315} (\bibinfo {year} {2016}{\natexlab{b}})}\BibitemShut
  {NoStop}%
\bibitem [{\citenamefont {Shi}\ \emph {et~al.}(2026)\citenamefont {Shi},
  \citenamefont {Bai}, \citenamefont {Zhang}, \citenamefont {Zhao},\ and\
  \citenamefont {Li}}]{Shi2026}%
  \BibitemOpen
  \bibfield  {author} {\bibinfo {author} {\bibfnamefont {F.}~\bibnamefont
  {Shi}}, \bibinfo {author} {\bibfnamefont {G.}~\bibnamefont {Bai}}, \bibinfo
  {author} {\bibfnamefont {X.}~\bibnamefont {Zhang}}, \bibinfo {author}
  {\bibfnamefont {Q.}~\bibnamefont {Zhao}},\ and\ \bibinfo {author}
  {\bibfnamefont {L.}~\bibnamefont {Li}},\ }\href
  {https://arxiv.org/abs/2608.12785} {\bibinfo {title} {A minimum-cardinality
  genuinely unextendible product basis in three qutrits}} (\bibinfo {year}
  {2026}),\ \Eprint {https://arxiv.org/abs/2608.12785} {arXiv:2608.12785
  [quant-ph]} \BibitemShut {NoStop}%
\end{thebibliography}
\end{document}